\documentclass[12pt]{article}
\usepackage{color}
\usepackage{amsmath,amssymb}  

\newtheorem{definition}{Definition}[section]

\newtheorem{proposition}[definition]{Proposition}

\newtheorem{theorem}[definition]{Theorem}

\newtheorem{lemma}[definition]{Lemma}

\newtheorem{rmk}{Remark}[section]

\numberwithin{equation}{section}

\newcommand{\nonu}{\nonumber \\}

\newcommand{\hs}[1]{\hspace{#1 mm}}

\def\cA{{\cal A}}   \def\cB{{\cal B}}   
   \def\cE{{\cal E}}   \def\cF{{\cal F}}

      \def\cU{{\cal U}}
\def\cV{{\cal V}}      
\def\cY{{\cal Y}}    

\def\fa{{\mathfrak a}}
\def\fb{{\mathfrak b}}
\def\fc{{\mathfrak c}}

\def\ff{{\mathfrak f}}
\def\fg{{\mathfrak g}}
\def\fh{{\mathfrak h}}
\def\fm{{\mathfrak m}}
\def\fn{{\mathfrak n}}

\newcommand{\BB}{\mbox{${\mathbb B}$}}
\newcommand{\CC}{{\mathbb C}}
\newcommand{\EE}{{\mathbb E}}
\newcommand{\II}{{\mathbb I}}

\newcommand{\RR}{\mbox{${\mathbb R}$}}

\newcommand{\ZZ}{{\mathbb Z}}

\newcommand{\wh}[1]{\widehat{#1}}
\newcommand{\wt}[1]{\widetilde{#1}}
\newcommand{\mb}[1]{\hs{4}\mbox{#1}\hs{4}}

\newcommand{\half}{\frac{1}{2}}

\newcommand{\prf}{\underline{Proof:}\ }
\newcommand{\finprf}{\null \hfill {\rule{5pt}{5pt}}\\[2.1ex]\indent}

\newcommand{\atopn}[2]{\genfrac{}{}{0pt}{}{#1}{#2}}

\definecolor{brique}{rgb}{.9,.2,0}
\definecolor{blvert}{rgb}{0,.8,.85}
\definecolor{vertcl}{rgb}{0,1,.7}
\newcommand\vertcl[1]{\textcolor{vertcl}{#1}}
\newcommand\blvert[1]{\textcolor{blvert}{#1}}
\newcommand\brique[1]{\textcolor{brique}{#1}}
\def\lapth{
\begin{picture}(136,70)(0,-15)\thicklines
\put(0,0){\vertcl{\rule{20pt}{4pt}}}
\put(19,1){\vertcl{\line(1,3){23}}} 
\put(20,1){\vertcl{\line(1,3){23}}} 
\put(21,1){\vertcl{\line(1,3){23}}}
\put(22,1){\vertcl{\line(1,3){23}}}
\put(45,70){\vertcl{\line(1,-3){23}}} 
\put(44,70){\vertcl{\line(1,-3){23}}} 
\put(43,70){\vertcl{\line(1,-3){23}}}
\put(42,70){\vertcl{\line(1,-3){23}}}
\put(2,24){\vertcl{\rule{120pt}{4pt}}}
\put(65,0){\vertcl{\rule{60pt}{4pt}}}
\put(5,37){\Huge{\brique{\textbf{L}}}} 
\put(62,37){\Huge{\brique{\textbf{PTh}}}}
\put(12,-8){\blvert{\rule{92pt}{3.5pt}}}
\put(24,-15){\blvert{\rule{57pt}{3.5pt}}}
\put(36,-22){\blvert{\rule{30pt}{3.5pt}}}
\end{picture}
\raisebox{35pt}{
\begin{minipage}{320pt}\begin{center}
\textbf{Laboratoire d'Annecy-le-Vieux de Physique
Th\'eorique}\\[4ex]
website: \texttt{http://lappweb.in2p3.fr/lapth-2005/}
\end{center}
\end{minipage}}\\
\vspace{10pt}\quad \hrulefill\\
\vspace{10pt}}

\newcommand{\beq}{\begin{equation}}
\newcommand{\eeq}{\end{equation}}
\newcommand{\ben}{\begin{eqnarray}}
\newcommand{\een}{\end{eqnarray}}

\begin{document}

\pagestyle{empty}
\setcounter{page}{0}
\hspace{-1cm}\lapth

\vfill\vfill
\begin{center}
{\LARGE  {\sffamily 
Nested Bethe ansatz for "all" closed spin chains}}

\vspace{10mm}
  
{\large S. Belliard\footnote{samuel.belliard@lapp.in2p3.fr} and E. 
Ragoucy\footnote{eric.ragoucy@lapp.in2p3.fr}
\\[.42cm]
 \textsl{Laboratoire de Physique Th{\'e}orique LAPTH\footnote{UMR 
5108 
du CNRS, associ{\'e}e {\`a} l'Universit{\'e} de
Savoie.}\\[.242cm]
 BP 110, F-74941  Annecy-le-Vieux Cedex, France. } }

\end{center}
\vfill\vfill
\begin{abstract}

We present in an unified and detailed way the Nested Bethe Ansatz for closed spin chains 
based on $\cY(gl(\fn))$, $\cY(gl(\fm|\fn))$,
$\wh\cU_{q}(gl(\fn))$ or $\wh\cU_{q}(gl(\fm|\fn))$ (super)algebras, with 
arbitrary representations (i.e. `spins') on each site of the chain. 
In particular, the case of indecomposable representations of 
superalgebras is studied.
The construction extends and unifies the results already obtained for spin chains 
based on $\cY(gl(\fn))$ or $\wh\cU_{q}(gl(\fn))$ and for some particular 
super-spin chains. We give the Bethe equations and 
  the form of the Bethe vectors. The case of $gl(2|1)$, 
$gl(2|2)$ and $gl(4|4)$ superalgebras (that are related to AdS/CFT 
correspondence) is also detailed.
\end{abstract}

\vfill

\begin{center}
\end{center}

\vfill

\rightline{LAPTH-1245/08}
\rightline{\texttt{arXiv:0804.2822 [math-ph]}}
\rightline{April 2008}

\newpage
\pagestyle{plain}
\markright{\today\dotfill DRAFT\dotfill }

\section{Introduction}

Finding eigenvectors and eigenvalues of a transfer matrix is a
fundamental problem in integrable systems. It started with the work of 
Bethe, which led to the celebrated Bethe ansatz \cite{Beth}. 
Then, the framework of the Quantum Inverse Scattering Problem 
 based on the Yang-Baxter equation became one of the most used way to 
adress this question. This technique is being developed since the $70$'s by 
the Leningrad 
School, see for example the review \cite{Fad} and references therein. 
Since then, numerous publications have been devoted to the subject, 
so that it is becoming difficult to make exhaustive citation. 
In the seek of such an impossible task, we will focus on Bethe equations 
and Bethe vectors for closed (or periodic) spin chains based 
on $gl(\fn)$ or $gl(\fm|\fn)$ algebras (leading to generalized XXX 
(super)spin chains) and their quantum 
deformations (leading to generalized XXZ (super)spin chains). 
The resolution of the general spin chain model started
with the calculation of the Bethe equations, computed for $gl(\fn)$ 
chains (with spins in the fundamental representation) in \cite{KuResh81,OW}.
Other cases (e.g. combining different spins) have been done in 
\cite{AbadRios}, see also \cite{KuReSk,byebye}.
Closed spin chains  based on $gl(\fm|\fn)$ superalgebras in the
distinguished diagram were
studied in \cite{Kulish} and \cite{saleur} and, in the case of alternating
fundamental-conjugate  
representations of $gl(\fm|\fn)$ in \cite{Mart1}. In 
\cite{Tsuboi}, closed spin chains in the fundamental representation but for 
any type of Dynkin diagram where studied using the Baxter 
$Q$-operator, and generalized in \cite{Tsuboi2} to a chain where all 
the spins are in a (type 1) typical representation depending on a free 
parameter. The general approach for arbitrary representations with 
any type of Dynkin diagram was done in \cite{RS}. 
General approach using Hirota equation was done in \cite{zaka}.
The case of quantum deformations was dealt in \cite{Jim,MezNep} for 
algebras (see also \cite{ACDFR2} for a global treatment)

However, in most of the above papers, one computes the Bethe equations and  
the transfer matrix 
eigenvalues, but not the Bethe vectors (i.e. the 
transfer matrix eigenvectors). To get them, one needs a more 
involved Bethe ansatz, the algebraic Bethe ansatz \cite{FadTak} (for 
rank 1 algebras) and its refinement to higher rank algebras, the 
nested Bethe ansatz \cite{KuSkly,KuResh83}. Algebraic Bethe ansatz for 
a general $gl(2)$ spin chain can be found in \cite{JMM-ollala}. 
Generalization to 
superalgebras has been done in some particular cases, such as the 
$gl(1|2)$ superalgebra \cite{Frank}. Nested Bethe ansatz for generalized 
XXZ spin chains with fundamental representations has been studied \cite{BdVV}.
Alternating generalized XXZ super-spin chain has been treated 
 in \cite{RibMart}.

More recently, a unified presentation for Bethe vectors of $gl(\fn)$ 
and $\cU_{q}(gl(\fn))$ spin chains has been developped 
\cite{MTV,TaVa}, 
producing a `trace formula' for Bethe vectors. This trace formula was 
shown to obey the same recursion formula that is obtained from the 
nested Bethe ansatz, proving equivalence between the two approaches. 

Let us also mention an alternative approach \cite{KhoPak,OPS} to the 
construction of Bethe 
vectors, using currents in the Drinfeld presentation of (quantum) 
algebras. The construction is 
off-shell (i.e. without any reference to Bethe equations) and thus may 
open a way to compute correlation functions. In this formalism, the 
construction is done without any reference to a highest weight, but 
rather computing modulo a suitably defined Borel subalgebra. The 
Bethe vectors are then viewed as special projections of currents that obey 
(Bethe ansatz) comultiplication properties. Note that these properties are 
valid even without 
Bethe equations: these equations appear when asking the off-shell Bethe 
vectors to be eigenvectors of the transfer matrix \cite{FKPR}.
The construction (and the connection 
with the previous approach) has been done for 
$\cY(\fn)$ and $\wh\cU_{q}(\fn)$ algebras \cite{KPT}. The case of 
(deformed) superalgebras remains to be treated in this formalism. 

\null

In the present paper we present in a unified way the nested Bethe 
ansatz for spin chains based on $gl(\fn)$, 
$gl(\fm|\fn)$,
$\cU_{q}(gl(\fn))$ and $\cU_{q}(gl(\fm|\fn))$ (super)algebras, with 
arbitrary representations (i.e. `spins') on each site of the chain.
In the case of (quantum) algebras, the construction is equivalent to 
the `trace formula' approach, and we make contact between the two 
presentations. Our construction also works for (quantum) superalgebras and we 
exhibit a `supertrace formula' for the Bethe vectors. The technique is essentially 
algebraic and works as soon as the representations on the spin chains 
are highest weight.
Then we use the Shapovalov form to prove 
orthogonality condition between Bethe vectors.

The plan of the paper is as follows. In section \ref{sec:algebras}, we 
introduce the different algebras that are concerned with our approach, 
presenting their $R$-matrices and their finite-dimensional 
irreducible representations. In section \ref{sec:ABA}, as a warm up, 
we remind the algebraic Bethe ansatz, which deals with 
spin chains based on $gl(2)$, $gl(1|1)$ algebras
and their quantum deformations. Then,  in section \ref{sec:NBA}, we 
perform the nested Bethe ansatz in a very detailled and pedestrian 
way\footnote{At least to our opinion\ldots} and up to the end. 
Finally, in section \ref{sec:betheV}, we study the Bethe vectors that 
have been constructed in the prevous section, showing connection with 
the `trace formula' and generalizing it to (quantum) superalgebras. Some 
examples of Bethe vectors are also given. The case of $gl(2|1)$, 
$gl(2|2)$ and $gl(4|4)$ superalgebras (that are related to AdS/CFT 
correspondence) is detailed in section \ref{sect:ex-super}. An 
appendix is devoted to the presentations of the finite dimensional 
(super)algebras used in the paper.

\section{Algebraic structures for closed spin chains\label{sec:algebras}}    

\subsection{Auxiliary graded spaces\label{sec:graded-aux}}
We use the so-called auxiliary space framework, a useful notation
 for the $R$-matrix formalism.
In this formalism, one deals with multiple tensor product of vectorial spaces 
$\cV \otimes \dots \otimes \cV$ and operators (defining an algebra 
$\cA$) therein.
For a matrix valued operator 
$A:= \sum_{ij} E_{ij} \otimes A_{ij} \in End(\cV) \otimes \cA$,
and any numbers $k\leq m$ we set:
\ben
A_k:=\sum_{ij} \II^{\otimes (k-1)}\otimes E_{ij} \otimes  \II^{\otimes (m-k)}
\otimes  A_{ij}  \in End(\cV^{\otimes m}) \otimes \cA\,,
\ 1 \leq k \leq m
\een
where $E_{ij}$ are elementary matrices with 1 at
position $(i,j)$ and 0 elsewhere.

The notation is also valid for complex matrices, taking $\cA:= \CC$ 
and using the isomorphism $End(\cV) \otimes \CC \sim End(\cV)$.

When $A \in End(\cV) \otimes End(\cV) \otimes \cA$, for $k,l$ such that 
$1 \leq k<l \leq m$, we denote by $A_{k l}$ the operator in $End( 
\cV^{\otimes m})\otimes \cA$ defined by
\ben
A_{k l}:= \sum_{ijab} \II^{\otimes (k-1)}\otimes E_{ij} \otimes  
\II^{\otimes (l-k-1)}
\otimes E_{ab} \otimes \II^{\otimes (m-l)}\otimes A_{ijab} \,.
\een

\null

We will work on $\ZZ_{2}$-graded spaces $\CC^{\fm\vert \fn}$.
The elementary $\CC^{\fm\vert \fn}$ column vectors $e_{i}$ (with 1 at 
position $i$ and 0 elsewhere) and elementary 
$End(\CC^{\fm\vert\fn})$ matrices $E_{ij}$ have grade 
\begin{equation}
[e_{i}]=[i] \mb{and} [E_{ij}]=[i]+[j]. 
\end{equation}
This grading is also extended to the superalgebras we deal with, see 
section \ref{sec:RTT} below.

The tensor product is graded accordingly:
\begin{equation}
(A_{ij} \otimes A_{kl})(A_{ab} \otimes A_{cd}) =
(-1)^{([k]+[l])([a]+[b])}(A_{ij}A_{ab}\otimes A_{kl}A_{cd})\,.
\end{equation}

To simplify the presentation we work with the
\textit{distinguished}
$\ZZ_{2}$-grade defined by
\begin{equation}
[i]=\left\{ \begin{array}{ll}  0\,, & 1 \leq i \leq \fm\,,\\
			       1\,, & \fm+1\leq i \leq \fm+\fn\,.
  \end{array}\right.
\end{equation}
Simplification in the expressions follows from
 the following rule:
\beq
[i][j]=[i] \mb{when}\, i\leq j\,,
\eeq
which is valid only for the distinguished grade. Generalization
 to other gradings is  easy to do. 

The non graded case is obtained setting formally $\fn=0$ in the above 
expressions.

\subsection{$R$ matrices}

In what follows, we will deal with different types of 
$R \in End(\cV) \otimes End(\cV) $ matrices, 
all obeying (graded) Yang-Baxter equation
 (writen in auxiliary space 
$End(\cV) \otimes End(\cV) \otimes End(\cV)$):
 \ben
R_{12}(u_{1},u_{2})\ R_{13}(u_{1},u_{3})\
R_{23}(u_{2},u_{3})&=&R_{23}(u_{2},u_{3})\ R_{13}(u_{1},u_{3})\
R_{12}(u_{1},u_{2})
  \label{YBE}
\een
and unitarity relation:
\ben
  R_{12}(u,v) R_{21}(v,u) &=&\zeta(u,v) \,\II \otimes \II\,,
  \label{Unit}
\een
where $\zeta(u,v)$ is a $\CC$-function depending on the model under 
consideration (see below).
These are the two fondamental properties used to construct 
transfer matrix for periodic spin chains. Below, we focus on infinite
dimensional associatives algebras based on $gl(\fn)$ and $gl(\fm|\fn)$ Lie (super) 
algebras: 
Yangians $\cY \big(gl(\fn)\big)\equiv \cY(\fn)$, super Yangians
$\cY \big( gl(\fm|\fn)\big)\equiv \cY(\fm|\fn)$, quantum affine (super) 
group  
$\cU_{q}(\wh{gl}(\fn)) \equiv \wh\cU_{q}(\fn)$ and 
$\cU_{q}(\wh{gl}(\fm|\fn))\equiv \wh\cU_{q}(\fm|\fn)$.
We note these algebras $\cA_{\fn}= Y(\fn)$ or $\wh\cU_{q}(\fn)$ and 
$\cA_{\fm|\fn}= \cY(\fm|\fn)$ or $\wh\cU_{q}(\fm|\fn)$.  As a
notation, we will write also $\cA_{\fm|0}=\cA_{\fm}$.

Depending on the choice 
of the algebra, we will construct different spin chains: 
\begin{itemize}
\item \textsl{\textbf{For $gl(\fn)$  or generalized XXX spin chains,}} the algebra 
is the Yangian 
$\cY(\fn)$ 
with rational $R$-matrix:
\ben
R_{12}(u,v) &=& R_{12}(u-v) =(u-v) \,\II \otimes \II - \hbar\,P_{12} 
\label{RR}\mb{with}
P_{12}=\sum_{i,j=1}^n E_{ij}\otimes E_{ji} \quad
\een
where $u$ is a spectral parameter 
over the field $\CC$ and
$P_{12}$ is the permutation operator 
($P_{12}(a\otimes b)=b \otimes a$). 
It is the simplest rational solution of the Yang-Baxter
equation found by Yang and Baxter in \cite{Baxt,Yg} and studied 
by Drinfel'd \cite{D2,D3} in connection with enveloping Lie 
algebras. When $\fn = 2$ and all the spins  are in 
fundamental (i.e. spin $\half$) representation, the spin chain model 
constructed from this $R$-matrix is the celebrated Heisenberg XXX model
\cite{He}.  

The unitarity relation reads:
\ben
R_{12}(u,v)\,R_{21}(v,u)&=&(u-v-\hbar)(v-u-\hbar)\,\II \otimes \II \,.
\label{URR}
\een
Note that the matrix is symmetric:
\begin{equation}
R_{21}(u)\equiv P_{12}\,R_{12}(u)\,P_{12}=R_{12}(u)\,.
\end{equation}
{From} the mathematical point of view, 
the parameter of deformation $\hbar$ is irrelevent 
since we have the isomorphism 
$\cY_{\hbar}(\fn)\sim \cY_{\hbar'}(\fn)$ for any non-vanishing values 
of $\hbar$ and $\hbar'$. For this reason, it is in general set to 1 in 
the mathematical litterature. However, in spin chains studies, it is 
set to $\pm i$, to ensure that the Hamiltonian is Hermitian. 
In this paper, we keep it free to encompass these two 
conventions.

\item \textsl{\textbf{For $gl(\fm|\fn)$ or supersymmetric XXX spin chains,}} 
one considers the super-Yangian $\cY(\fm|\fn)$, 
introduced in \cite{KuSkly,Kulish,N} with the same form (\ref{RR}) for the 
$R$-matrix and unitarity relation (\ref{URR}), but with
 a  $\ZZ_2$ graded auxiliary space.
The permutation operator in the graded space takes the form:
\begin{equation}
P_{12} = \sum_{i,j=1}^{\fm+\fn} (-1)^{[j]}E_{ij}\otimes E_{ji} \,,
\end{equation}
so that we have $P_{12}(a\otimes b)=(-1)^{[a][b]}\,b \otimes a$.
 
\item \textsl{\textbf{We will also deal with $\wh\cU_{q}(\fn)$ or 
generalized XXZ spin chains.}}
 In that case, one considers the $R$-matrix of the (centerless) affine 
 quantized algebra $\wh\cU_{q}(\fn)$:
\ben
R_{12}(u,v) &=& R_{12}(\frac{u}{v})=
(\frac{u\,q}{v}-\frac{v}{u\,q})\sum_{a=1}^{{\fn}} 
E_{aa}\otimes E_{aa} 
+ (\frac{u}{v}-\frac{v}{u} )\sum_{1\leq a\neq b \leq{\fn}} 
E_{aa} \otimes E_{bb}  \nonumber \\
 && + (q-q^{-1}) \sum_{1\leq a\neq b \leq{\fn}} 
  (\frac{u}{v})^{sign(b-a)} E_{ab}\otimes E_{ba}
\label{RT}
\een
with unitarity relation:
\ben
R_{12}(u,v)\,R_{21}(v,u)&=&
(\frac{u\,q}{v}-\frac{v}{u\,q})(\frac{v\,q}{u}-\frac{u}{v\,q}) \,
\II \otimes \II
\een
where $q$ is a generic complex number, not root of unity.
It has been introduced by Jimbo or Faddeev, Reshetikhin and 
Takhtajan \cite{Jim,FRT}. When $\fn = 2$ and the spins lie in 
fundamental representation we recover the Heisenberg XXZ model.

\item \textsl{\textbf{The last cases considered are 
$ \wh\cU_{q}(\fm|\fn)$ or supersymmetric XXZ spin chains.}}
 The $R$-matrix of the (centerless) affine 
 quantized algebra $\wh\cU_{q}(\fm|\fn)$ 
reads \cite{PerSch,BazSh,Z3}:
\ben
R_{12}(u,v) &=& R_{12}(\frac{u}{v})=
\sum_{a=1}^{\fm+\fn} 
(\frac{u}{v}q^{1-2[a]}-\frac{v}{u}q^{-1+2[a]})
E_{aa}\otimes E_{aa}
+(\frac{u}{v}-\frac{v}{u} )\sum_{1\leq a\neq b \leq{\fm+\fn}} 
E_{aa} \otimes E_{bb} \nonumber \\
 &&+ (q-q^{-1}) \sum_{1\leq a\neq b \leq{\fm+\fn}} 
   (\frac{u}{v})^{sign(b-a)} (-1)^{[b]} E_{ab}\otimes E_{ba}\,.
\label{RTS}
\een
The auxiliary space is graded, and
the unitarity relation reads:
\ben
R_{12}(u,v)\,R_{21}(v,u)&=&(\frac{u\,q}{v}-\frac{v}{u\,q})
(\frac{v\,q}{u}-\frac{u}{v\,q}) \,\II \otimes \II\,.
\een

\item \textsl{\textbf{We will encompass all these cases writing:}}
\ben
R_{12}(u,v) &=&\sum_{a=1}^{{\fm+\fn}} {\mathfrak a_a}(u,v)  E_{aa}
\otimes E_{aa}+ {\mathfrak b}(u,v) \sum_{1\leq a\neq b \leq{\fm+\fn}}E_{aa} 
\otimes E_{bb} \nonu
&&+ \sum_{1\leq a\neq b \leq{\fm+\fn}} {\mathfrak c_{ab}(u,v)} 
E_{ab}\otimes E_{ba}
\label{RU}
\een
with the following identifications:
\begin{equation}
\mb{For}Y(\fn):\qquad
{\mathfrak a_a}(u,v)=u-v-\hbar \mb{;}
{\mathfrak b}(u,v)= u-v\mb{;}
{\mathfrak c_{ab}(u,v)} = -\hbar 
\label{eq:defac-Y}
\end{equation}
\begin{equation}
\begin{array}{ll}
\mb{For}Y(\fm|\fn):\qquad
&{\mathfrak a_a}(u,v)=u-v-(-1)^{[a]}\,\hbar
\mb{;}{\mathfrak b}(u,v)= u-v \\[1.ex]
&{\mathfrak c_{ab}(u,v)} = -(-1)^{[b]}\,\hbar 
\end{array}
\end{equation}
\begin{equation}
\begin{array}{ll}
 \mb{For}\wh\cU_{q}(\fn):\qquad
&\displaystyle{\mathfrak a_a}(u,v)=(\frac{u\,q}{v}-\frac{v}{u\,q})
\mb{;}\displaystyle{\mathfrak b}(u,v)=\frac{u}{v}-\frac{v}{u} \\[1.2ex]
&\displaystyle
{\mathfrak c_{ab}}(u,v) = (q-q^{-1})(\frac{u}{v})^{sign(b-a)} 
\end{array}
\end{equation}
\begin{equation}
\begin{array}{ll}
\mb{For}\wh\cU_{q}(\fm|\fn):\qquad
&\displaystyle{\mathfrak a_a}(u,v)
=(\frac{u}{v}q^{1-2[a]}-\frac{v}{u}q^{-1+2[a]})
\mb{;}\displaystyle{\mathfrak b}(u,v)=\frac{u}{v}-\frac{v}{u}\\[1.2ex]
&\displaystyle{\mathfrak c_{ab}}(u,v) = 
(q-q^{-1})(\frac{u}{v})^{sign(b-a)}(-1)^{[b]} 
\end{array}
\label{eq:defac-Uq}
\end{equation}
In this notation, the unitary relation reads
\begin{equation}
\zeta(u,v)=\fa_{1}(u,v)\fa_{1}(v,u)\,.
\end{equation}
Remark that we have the properties
\begin{eqnarray}
&&\fa_{k}(u,v)\,\fa_{k}(v,u) = \fa_{l}(u,v)\,\fa_{l}(v,u)\,,\ \forall\ 
k,l\\
&&\fb(u,v) = -\fb(v,u) \mb{and}
\fc_{ab}(u,v) = (-1)^{[a]+[b]}\ \fc_{ba}(v,u)
\end{eqnarray}
\end{itemize}

We will also use `reduced' $R$-matrices $R^{(k)}(u)$, deduced from 
$R(u)$ by suppressing all the terms containing indices $j$ with $j < k$:
\ben
R_{12}^{(k)}(u,v) &=&\sum_{a=k}^{{ \fm+\fn}} {\fa_a}(u,v)
E_{aa}\otimes E_{aa}+ \fb(u,v) \sum_{k \leq a\neq b \leq \fm+\fn}
E_{aa} \otimes E_{bb}\nonu
&&+ \sum_{k\leq a\neq b \leq \fm+\fn} \fc_{ab}(u,v) 
E_{ab}\otimes E_{ba}\,.
\een

Hence, we have $R_{12}^{(1)}(u,v)=R_{12}(u,v)$, and more generally 
$R_{12}^{(k)}(u,v)$ 
corresponds to the embbedding $\cA_{\fm+1-k |\fn}\subset \cA_{\fm|\fn}$ 
when 
$k \leq \fm+1$ 
or $\cA_{0|\fn-(k-\fm-1)}\subset \cA_{\fm|\fn}$ otherwise. In the 
following, to make the presentation concise, we will write, for a generic $k$, 
$\cA_{\fm+1-k|\fn}$, keeping 
in mind that one should write $\cA_{0|\fn-(k-\fm-1)}$ when $k>\fm+1$.

We define the normalized reduced $R$-matrices 
\begin{equation}
\RR_{12}^{(k)}(u,v)=\frac{1}{\fa_{k}(u,v)}\,R_{12}^{(k)}(u,v) 
\mb{such that} \RR^{(k)}_{12}(u,v)\,\RR^{(k)}_{21}(v,u)
=\II\otimes\II\,.
\label{Runit}
\end{equation}

\subsection{RTT relation and transfer matrix\label{sec:RTT}}
 
The algebraic structures associated to spin chains
are defined using the RTT relations \cite{D3,FRT}. They allow to 
generate all the relations between each generator
of the graded unital associative algebra $\cA_{\fm|\fn}$.
We gather the $\cA_{\fm|\fn}$ generators  into a 
$(\fm+\fn) \times (\fm+\fn)$  matrix 
acting in an auxiliary space $\cV=\CC^{\fm|\fn}$ whose entries 
are formal series of a complex parameter $u$, 
\ben
T(u)= \sum_{i,j=1}^{\fm+\fn}  E_{ij} \otimes t_{ij}(u)
\in  \cV \otimes \cA[[u,u^{-1}]] \,. \nonumber
\een
Since the auxiliary space $End(\CC^{\fm|\fn})$ is interpreted as a 
representation of $\cA_{\fm|\fn}$ (see below), 
the $\ZZ_{2}$-grading of $\cA_{\fm|\fn}$ must correspond to the one 
defined on $End(\CC^{\fm|\fn})$ matrices (section \ref{sec:graded-aux}). 
Hence, the generator $t_{ij}(u)$ has grade $[i]+[j]$, so that the 
monodromy matrix $T(u)$ is globally even. As for matrices, the tensor product 
of algebras will be graded, as well as between algebras and matrices, 
e.g.
\begin{eqnarray}
\big(E_{ij} \otimes t_{ij}(u)\big)\,\big(E_{kl} \otimes t_{kl}(u)\big)=
(-1)^{([i]+[j])([k]+[l])}\,E_{ij}\,E_{kl} 
\otimes t_{ij}(u)\,t_{kl}(u)\,.
\end{eqnarray}

The `true' generators $t_{ij}^{(n)}$ of $\cA_{\fm|\fn}$ appear upon expansion of 
$t_{ij}(u)$ in $u$. For the (super) Yangians $\cY(\fn)$ and $\cY(\fm|\fn)$, 
$t_{ij}(u)$ is a series in $u^{-1}$:
\begin{equation}
t_{ij}(u)=\sum_{n=0}^\infty t_{ij}^{(n)} u^{-n} \mb{with}
t_{ij}^{(0)}=\delta_{ij}\,.
\end{equation}

In the quantum affine (super)algebra case, a complete description of
the algebras requires the introduction of two matrices $L^\pm(u)$ 
\ben
L^\pm(u)&=&\sum_{i,j=1}^{\fm+\fn}  E_{ij} \otimes L^\pm_{ij}(u) 
=\sum_{i,j=1}^{\fm+\fn}  E_{ij} \otimes \sum_{n=0}^\infty 
L^{\pm (n)}_{ij} u^{\pm n}
\een
with  relations: 
\ben
L^{+ (0)}_{ii}L^{- (0)}_{ii}=1\,,\,\forall\,i &\mbox{and}&
L^{+(0)}_{ij}=0=L^{- (0)}_{ji}\,,\ i<j\\
R_{12}(u,v)\ L^\pm_{1}(u)\ L^\pm_{2}(v) 
&=&L^\pm_{2}(v)\ L^\pm_{1}(u)\ R_{12}(u,v)\,,\\
R_{12}(u,v)\ L^\mp_{1}(u)\ L^\pm_{2}(v)
&=&L^\pm_{2}(v)\ L^\mp_{1}(u)\ R_{12}(u,v)\,.
  \label{RLL}
\een
However, in the context of evaluation representations it is sufficient 
to consider only one, say $T(u)=L^+(u)-L^-(u)$, to 
construct a transfer matrix, see e.g. \cite{ACDFR2} for more details.

Then, the RTT relations take the form:
\beq
  R_{12}(u,v)\ T_{1}(u)\ T_{2}(v)=T_{2}(v)\ T_{1}(u)\ R_{12}(u,v)\,. 
  \label{RTT}
\eeq
{From} the  $R$-matrix (\ref{RU}), we get the commutation relations through an 
expansion  on the graded basis $E_{ij}\otimes E_{kl}$:
\ben
\fb(u,v)\,\left[t_{ij}(u)\,,\,t_{kl}(v)\right\} &=&
\delta_{ik}\,\big(\fb(u,v)-\fa_i(u,v)\big)\,t_{kj}(u)\,t_{il}(v) 
\nonumber \\[1.2ex]
&&-(1-\delta_{ik})\,(-1)^{([i]+[k])\,([k]+[l])}\,\fc_{ik}(u,v)\,t_{kj}(u)\,t_{il}(v) 
 \nonumber \\[1.2ex]
&& -\delta_{jl}\,(-1)^{([k]+[l])([i]+[j])}\,\big(\fb(u,v)-\fa_j(u,v)\big)
\,t_{kj}(v)\,t_{il}(u) \nonumber \\[1.2ex]
&&+(1-\delta_{jl})\,(-1)^{[j]+[i][k]+[l]\,([i]+[k])}\,\fc_{lj}(u,v)
\,t_{kj}(v)\,t_{il}(u) 
\label{eq:relcom}
\een
with
\ben
\left[t_{ij}(u)\,,\,t_{kl}(v)\right\} &=& 
t_{ij}(u)\,t_{kl}(v)-(-1)^{([i]+[j])([k]+[l])}\,t_{kl}(v)\,t_{ij}(u)\\
&=& -(-1)^{([i]+[j])([k]+[l])}\,\left[t_{kl}(v)\,,\,t_{ij}(u)\right\}
\,.
\een

In the context of spin chain models, $T(u)$ is called the (algebraic) 
monodromy matrix. The connection with usual monodromy matrix is done 
upon representation (see next section).

$\cA$ has a Hopf algebra structure, whose 
  coproduct  $\Delta$ reads:
\ben
\Delta:\ \begin{array}{ccl}  
End(\CC^{\fm|\fn}) \otimes \cA_{\fm|\fn} & \to & End(\CC^{\fm|\fn})
\otimes \cA_{\fm|\fn} \otimes \cA_{\fm|\fn}  \\
 T(u) & \mapsto & \displaystyle
T(u)\dot{\otimes}T(u) = \sum_{i,j,k=1}^{\fm+\fn} (-1)^{([k]+[i])([k]+[j])}\,
E_{ij}\otimes t_{ik}(u)\otimes t_{kj}(u)
\end{array}\ 
\een 

More generally, one defines recursively for $L\geq 2$
\begin{equation}
\Delta^{(L+1)}=(\II^{\otimes (L-1)}\otimes \Delta)\circ\Delta^{(L)}\
:\ {\cA_{\fm|\fn}} \to {\cA_{\fm|\fn}}^{\otimes (L+1)}
\end{equation}
with $\Delta^{(2)}=\Delta$ and $\Delta^{(1)}=\II$. The map 
$\Delta^{(L)}$ is an algebra homomorphism.

One defines the transfer matrix as the  supertrace over the 
auxiliary space of
the monodromy matrix:
\beq
t(u) = str\big(T(u)\big) = \sum_{i=1}^{\fm+\fn}(-1)^{[i]}t_{ii}(u)
\,.
\label{stm}
\eeq
Relations (\ref{RTT}) and (\ref{Unit}) then show that the transfer 
matrices at two different values of the spectral parameter commute
\beq
[t(u),t(v)] = 0 \,.
\label{invol}
\eeq

Thus, $t(u)$ generates (via an expansion in $u$) a set of $L$ (the 
number of sites)
independent integrals of motion or charges in involution which ensure 
integrability of the model. 
  
The diagonalisation of the transfer matrix can be 
done in an algebraic way when working in a highest weight 
representation. Thus, we briefly describe the representation 
theory of the algebras we use. 
 
\subsection{Finite dimensional representations of $\cA_{\fm|\fn}$ 
and spin chains}

The fundamental point in using the ABA is to know a pseudo-vacuum 
for the model. In the mathematical framework 
it is equivalent to know a highest weight vector for 
the representation of the algebra which underlies the model. 
We describe the link between highest  weight vector of the standard 
finite dimensional Lie (super)algebras $gl(\fn)$ or 
$gl(\fm|\fn)$  and the infinite dimensional (graded) algebras
$\cA_{\fn}$ or $\cA_{\fm|\fn}$.

\begin{definition}
A representation  of $\cA_{\fm|\fn}$ is called \textit{highest weight} 
if there exists a nonzero vector $\Omega$ 
such that:
\begin{equation}
t_{ii} (u)\, \Omega = \Lambda_{i}(u)\, \Omega \mb{and}
t_{ij} (u)\, \Omega = 0  \text{ if } i>j   
\label{higvectU}
\end{equation}  
for some scalars $\Lambda_{i}(u)$ $\in$ $\CC$ $[[u^{-1}]]$.
$\Lambda(u)= (\Lambda_{1}(u),\dots,\Lambda_{\fm+\fn}(u))$ is 
called 
the highest weight and $\Omega$ the highest weight vector.
\end{definition}

The action of the T-matrix on $\Omega$ gives a triangular matrix. We
can interpret the operators $t_{ij} (u)$ for $i \neq j$ as creation
or annihilation  operators.
The main theorem on highest weights is:
\begin{theorem}
Every finite-dimensional irreducible representation of $\cA_{\fn}$ or 
$\cA_{\fm|\fn}$ is  highest weight. 
Moreover, it contains a unique 
 (up to scalar multiples) highest weight vector.
\end{theorem}
This theorem is presented in \cite{D2} for $\cY(\fn)$,  \cite{Z1} 
for
$\cY(\fm|\fn)$,  \cite{Rosso} for $\wh\cU_q(n)$ and \cite{Z2} for 
$\wh\cU_q(\fm|\fn)$.

\null

To construct such representations, one uses the evaluation morphism, 
which relates the infinite dimensional algebra $\cA_{\fm|\fn}$ to its 
finite dimensional subalgebra $\cB_{\fm|\fn}$. The correspondence 
between the algebras $\cA$ and $\cB$ is given in table (\ref{tableAB}). The 
algebraic structure of the $\cB_{\fm|\fn}$ algebras and their 
irreducible finite dimensional representations are described in 
the appendix.
\begin{equation}  
\begin{array}{|c|c|c|c|c|}
\hline\displaystyle
\rule{0ex}{3.2ex}
\mbox{(Super)algebra } \cA_{\fm|\fn} & \cY(\fn) & \cY(\fm|\fn) 
& \wh\cU_{q}(\fn) & \wh\cU_{q}(\fm|\fn)
\\[1.2ex]
\hline
 \rule{0ex}{2.6ex}
\mbox{Subalgebra }\cB_{\fm|\fn} & gl(\fn) & gl(\fm|\fn) 
& \cU_{q}(\fn) & \cU_{q}(\fm|\fn)
\\[1.2ex]
\hline
\end{array}
\label{tableAB}
\end{equation}  
The evaluation morphism with parameter $a\in\CC$ is given by
\begin{equation}
ev_{a}:\ \left\{\begin{array}{lcl} 
\cV\otimes\cA & \to & \cV\otimes\cB \\[1.2ex]
T(u) &\mapsto & 
\begin{cases}
\displaystyle (u-a)\II -\hbar\,\EE \mb{for} \cY(\fm|\fn) \\[2.1ex]
 \displaystyle  \frac{u}{a}\, L^+ - \frac{a}{u}\,L^- \mb{for} \wh\cU_q(\fm|\fn)
\end{cases}
\end{array}\right.
\end{equation}
where 
\ben
\EE &=& \sum_{i,j=1}^{\fm+\fn} (-1)^{[j]}\,E_{ij}\otimes\cE_{ji}  
\in \cV\otimes gl(\fm|\fn)
\mb{and}
L^\pm =  \sum_{i,j=1}^{\fm+\fn} E_{ij}\otimes l^\pm_{ij} 
\in \cV\otimes\cU_q(\fm|\fn)\quad
\een
with the convention $gl(\fm|0)\equiv gl(\fm)$ and 
$\cU_q(\fm|0)\equiv \cU_q(\fm)$ as for the infinite dimensional 
superalgebras $\cY(\fm|\fn)$ and $\wh\cU_q(\fm|\fn)$.

{From} the evaluation morphism $ev_{a}$ and a highest weight representation 
$\pi_{\lambda}$ of $\cB$, one can construct a highest weight 
representation of $\cA$, 
called evaluation representation:
\begin{equation}
\rho^\lambda_{a}=ev_{a}\,\circ\,\pi_{\lambda}\ :\ \cA_{\fm|\fn}\ 
\stackrel{ev_{a}}{\longrightarrow}\ \cB_{\fm|\fn}\  
\stackrel{\pi_{\lambda}}{\longrightarrow}\ \cV_{\lambda}\,.  
\end{equation}
The weight of this evaluation representation is given by
$\Lambda(u)=\big(\Lambda_{1}(u),\ldots,\Lambda_{\fm+\fn}(u)\big)$, with
\begin{equation}
\Lambda_{j}(u) =\begin{cases}
 u-a-(-1)^{[j]}\,\hbar\,\lambda_j \mb{for} \cY(\fm|\fn) 
\\[1.2ex]
\displaystyle
(-1)^{[j]}\,\left(\frac{u}{a}\,\eta_{j}q^{\lambda_{j}} - 
\frac{a}{u}\,\eta_{j}q^{-\lambda_{j}}\right)
\mb{for}\wh\cU_q(\fm|\fn)
\end{cases}
\ j=1,\ldots,\fm+\fn
\label{eq:Lambda-eval}
\end{equation}
where $\lambda_{j}$, $j=1,\ldots,\fm+\fn$ are the weights of the $\cB_{\fm|\fn}$ 
representation (see appendix).

More generally, one constructs tensor product of evaluation 
representations using the coproduct of $\cA$:
\ben
\Big(\otimes_{i=1}^{L}\,\rho^{\lambda^{\langle i\rangle}}_{a_i}\Big)\, \circ 
\Delta^{(L)}\Big(T(u)\Big) = \rho^{\lambda^{\langle 1\rangle}}_{a_{1}}\Big(T(u)\Big)\dot{\otimes}\,
\rho^{\lambda^{\langle 2\rangle}}_{a_{2}}\Big(T(u)\Big) \dot{\otimes}\cdots 
\dot{\otimes}\rho^{\lambda^{\langle L\rangle}}_{a_{L}}\Big(T(u)\Big)
\label{eq:mono-repr}
\een
where $\lambda^{\langle i\rangle}=(\lambda^{\langle i\rangle}_{1},\ldots,
\lambda^{\langle i\rangle}_{\fm+\fn})$, 
$i=1,\ldots,L$ are the weights of the $\cB_{\fm|\fn}$ representations. 
This provides a $\cA_{\fm|\fn}$ representation with weight
\ben
\Lambda_j(u)=\prod_{i=1}^{L}\Lambda^{\langle i\rangle}_j(u)
\,,\qquad j=1,\ldots,\fm+\fn
\label{VP}
\een
where $\Lambda^{\langle i\rangle}_j(u)$ have the form (\ref{eq:Lambda-eval}).

\null

In a spin chain context, the number $L$ of evaluation representations is the number of sites 
of the chain, the weights 
$\lambda^{\langle i\rangle}=(\lambda^{\langle i\rangle}_{1},\ldots,
\lambda^{\langle i\rangle}_{\fm+\fn})$, 
$i=1,\ldots,L$ characterize the $\cB$ representation (the spin) on 
each of these sites, and the evaluation parameter $a_{i}$ is the 
so-called inhomogeneity parameter at site $i$.

{From} the mathematical point of view, evaluation representations are 
relevant because of:
\begin{theorem}
All finite dimensional irreducible representations of $\cA_{\fm|\fn}$ 
can be constructed as (subquotient of) tensor products of evaluation 
representations.
\end{theorem}
This theorem is proven in \cite{Tara,CPeval} for $\cY(\fn)$ (see also 
\cite{Chari, Molirr}). It is proven in \cite{Bab,Jim86,Lusztig,Chari} for 
$\wh\cU_{q}(gl(\fn))$ and in \cite{Z1} for $\cY(\fm|\fn)$ (see also 
\cite{BR}). We don't know any reference for the case of 
$\wh\cU_{q}(gl(\fm|\fn))$, but the proof should be similar to the 
other cases, and, at least, one can construct a wide set of
finite dimensional 
irreducible representations from tensor product of evaluation 
representations.

Hence, the study of spin chains amounts to study finite 
dimensional representations of $\cA_{\fm|\fn}$, and the nested Bethe 
ansatz can be viewed as the construction of a Gelfand-Tsetlin type 
basis.

\subsection{Case of indecomposable (superalgebras) representations
\label{sect:indec}}
It is well-known that (most of the) Lie superalgebras (and 
specifically the $gl(\fm|\fn)$ superalgebras studied in the present paper)
 contain finite 
dimensional representations which are indecomposable. To discuss 
these special cases, we first remind some 
definitions about representations of Lie (super)algebras (see e.g. 
\cite{dico} for more details).

\subsubsection{Definitions}
We focus on finite dimensional representations.

\begin{definition}
A representation is called \textbf{irreducible} if it does not contain any 
non-trivial invariant subspace. A representation which is not 
irreducible is called \textbf{reducible}.
\end{definition}

\begin{definition}
A representation is called \textbf{fully reducible} if, for any invariant 
subspace, there exists a complementary subspace which is also 
invariant. A reducible representation which is not 
fully reducible is called \textbf{indecomposable}.
\end{definition}

It may be useful to illustrate these various definitions.
If one considers finite dimensional representations, the 
representation of the  Lie (super)algebra generators 
 are square matrices. Considering a general 
linear combination of all these matrices, we have roughly the 
following (very sketchy) picture
\begin{eqnarray}
\pi_{V}(\cA) = \left(\begin{array}{ccc} 
* & \ldots  & * \\
\vdots & * & \vdots \\
* & \ldots &  * \end{array}\right) && \mb{$V$ irreducible} \\
\pi_{V}(\cA) = \left(\begin{array}{cccc} 
* & * & 0 & 0 \\
* & * & 0 & 0 \\
0 & 0 & * & * \\
0 & 0 & * & * \end{array}\right) && \mb{$V$ fully reducible 
($V=V_{1}\oplus V_{2}$)} \\
\pi_{V}(\cA) = \left(\begin{array}{cccc} 
* & * & * & * \\
* & * & * & * \\
0 & 0 & * & * \\
0 & 0 & * & * \end{array}\right) && \mb{$V$ indecomposable} 
\end{eqnarray}
where $*$ denotes the non-zero entries and $V$ is the representation 
of $\cA$ under consideration.

\begin{theorem}
Any finite dimensional representation of semi-simple Lie algebras is fully 
reducible.
This is not the case of $gl(\fm|\fn)$ superalgebras. In particular, 
for these superalgebras, the tensor product of irreducible 
representations is not always fully reducible.
\end{theorem}
Examples of indecomposable representations and of indecomposable 
tensor products can be found for $sl(1|2)$ in e.g. \cite{vlad,marcu}.

\begin{definition}
A vector $v$ in a vector space $V$, representation of $\cA$, is called 
\textbf{cyclic} if the (iterative)
action of all the generators of $\cA$ on $v$ span all $V$.
\end{definition}
For example, in an irreducible highest weight representation, the 
highest weight vector is cyclic. This is not true for a fully 
reducible highest weight representation, which contains several 
highest weight vectors, all of them being needed to span the full representation. 
However, in an indecomposable representation (finite dimensional), 
there exists a highest weight vector which is cyclic (see example in 
\cite{vlad}).

\subsubsection{Application to spin super-chains}
It is natural to wonder whether the presence of indecomposable 
representations on a spin chain alters the Bethe ansatz technics. 
We argue (and prove in some cases)  that the algebraic Bethe ansatz still 
works. The 
reasonning is done for the $gl(\fm|\fn)$ superalgebras, but  
quite likely it applies to the deformed case.

The basic ingredient for the algebraic Bethe ansatz is the existence 
of a cyclic highest weight vector. Through the recursive application 
of creation-like generators (see section below), one constructs the 
eigenvectors of the transfer matrix. Hence, it is not the reducibility 
of the representations, but rather the existence of highest weight 
that guarantees the Bethe ansatz technics. Indeed, when the 
representations are fully reducible, one gets several highest weight 
vectors. In that case, 
 one needs to apply the ansatz on each of the highest weight 
vectors, but the technics is still valid. For indecomposable 
representations, since there is a cyclic highest weight vector,
it is very plausible that the Bethe ansatz works. 

In particular, for 
indecomposable representations obtained from the tensor product of 
two irreducible representations, one can prove that the Bethe ansatz 
indeed works in the following way. From the original spin chain 
(that contains 
the indecomposable representation(s)), one constructs a new 
 chain, where each of the sites carrying an indecomposable 
representation is replaced by two sites, one for each irreducible 
representation underlying the indecomposable one. 
Obviously, the new chain is equivalent to the original one. Moreover, 
since the new chain contains only irreducible representations, it is 
clear that one can apply the algebraic Bethe ansatz, to this chain as 
well as to the original one.

Since no classification of indecomposable representations is known, 
it is difficult to tell whether they can all be obtained from tensor 
products of irreducible ones. Nevertheless, we have argued above that 
the Bethe ansatz should still work in all cases. 

As a last remark, let us add that the algebraic structures 
underlying spin chains are not the finite dimensional 
(super)algebras, but rather the infinite dimensional ones 
(super-Yangians or affine quantum algebras). For these algebras, 
tensor products of representations are in most of the cases also 
irreducible.
Then, the spin chain as 
a whole appears as a sole (irreducible) representation of these 
algebras, although it is reducible for the finite dimensional algebra. 
Thus, it is natural to expect that indecomposable representations of 
the finite dimensioanl superalgebras appear as `usual' representations 
for the infinite dimensional one.

\section{Algebraic Bethe Ansatz for the case $\fm+\fn$=2\label{sec:ABA}}

In this section, we remind the framework of the Algebraic Bethe Ansatz 
(ABA) \cite{STF}
introduced  in order to compute 
eigenvalues and
 eigenvectors of the transfer matrix. 

For $\fm+\fn=2$, one can 
consider two different 
algebras: $\cA_2\equiv\cA_{0|2}\equiv\cA_{2|0}$ or $\cA_{1|1}$.
We write the monodromy matrix  in the following
matricial form:
\beq
T(u) = \left (\begin{array}{ccc} 
t_{11}(u) &&  t_{12}(u) \\  t_{21}(u) && t_{22}(u)
\end{array} \right )\,,
\eeq
and the transfer matrix as $t(u) = str(T(u)) = 
(-1)^{[1]}t_{11}(u) +(-1)^{[2]} t_{22}(u)$.  
Let $\Omega$ be the pseudo-vacuum state presented in previous section:
\ben
t_{11}(u) \, \Omega  &=& \Lambda_1(u) \,\Omega \mb{;}
t_{22}(u) \, \Omega  \,=\, \Lambda_2(u) \,\Omega \mb{;}
t_{21}(u) \, \Omega  \,=\,  0 \,.
\een

Using the ternary (RTT) relation one can find the following relations 
between the different
operators of $\cA_2$ or  $\cA_{1|1}$:
\ben
 t_{12}(u) t_{12}(v)&=& 
\begin{cases}  
 t_{12}(v) t_{12}(u), \mbox{ for $\cA_2$} \\[1.2ex]
 \fh(u,v) t_{12}(v)t_{12}(u), \mbox{ for $\cA_{1|1}$} 
\end{cases} \\
t_{11}(u)\,t_{12}(v) &=& \ff_{1}(u,v)\,t_{12}(v)\,t_{11}(u) + 
\fg^{+}_{1}(u,v)\,t_{12}(u)\,t_{11}(v)
\label{ExchAB} \\
t_{22}(u)\,t_{12}(v) &=& \ff_2(v,u)\,t_{12}(v)\, t_{22}(u) 
+\fg^{-}_{2}(v,u)\,t_{12}(u)\,t_{22}(v)
\label{ExchDB}
\een
where we have used the functions
\ben
\ff_i(u,v) &=& \frac{\fa_i(v,u)}{\fb(v,u)} \mb{;}
\fg^{\pm}_{i}(u,v)= \frac{\fc_{i \pm 1,i}(u,v)}{\fb(u,v)} 
\mb{;}
\fh(u,v)=(-1)^{[1]+[2]}\,\frac{\fa_{2}(u,v)}{\fa_{1}(u,v)}
\qquad
\een

Applying $M$ creation operators we 
generate a Bethe vector:
\beq
\Phi(\{u\}) \, = t_{12}(u_1) \dots t_{12}(u_M) \,\Omega.
\eeq

Demanding $\Phi(\{u\})$ to be an eigenvector of $t(u)$ leads
to a set of algebraic relations on the parameters 
$u_1,\dots,u_M$, the so-called Bethe equations.

The relation between creation operators prove the invariance (up to a 
function for $\cA_{1|1}$) of the
Bethe vector under reordering of the operators $t_{12}(u_{j})$. This
condition is usefull to compute the unwanted terms from the 
action of $t(u)$ on  $\Phi(\{u\})$. 
First,  we compute the action of 
$t_{11}(u)$ on $\Phi(\{u\})$:
\ben
t_{11}(u)\,t_{12}(u_1) \dots t_{12}(u_M) \,\Omega &=&  
\prod_{k=1}^{M} \ff_1(u,u_k)\ 
\Lambda_1(u) \,t_{12}(u_1) \dots t_{12}(u_M) \,\Omega
 \nonumber \\
&&+\, \sum_{k=1}^{M} \, P_k(u;\{u_{j}\}) \,t_{12}(u_1) \dots 
t_{12}(u_k\to u) \dots
t_{12}(u_M)\,\Omega\quad\\
P_k(u;\{u_{j}\}) &=& \fg^{+}_{1}(u,u_k)\,\prod_{j\neq k}^{M} 
\ff_1(u_k,u_j)\,  
\Lambda_1(u_k) \,.
\label{ABBO}
\een
where the notation $t_{12}(u_k\to u)$ is used to indicate 
the position of $t_{12}(u)$ in the ordered product.
$P_k(u;\{u_{j}\}) $ corresponds to the $(2^M-1)$ terms containing 
$\Lambda_1(u_k)$.
 The form of $P_1(u;\{u_{j}\}) $ is easily computed.
The other polynomials $P_k(u;\{u_{j}\}) $ are then computed using the 
commutation relation between the operators $t_{12}(u_{j})$ 
 and puting $t_{12}(u_k)$ on the left.
With the same method we compute the action of $t_{22}(u)$ on 
$\Phi(\{u\})$:
\ben
 t_{22}(u)\,t_{12}(u_1) \dots t_{12}(u_M)\, \Omega &=& \prod_{k=1}^{M} 
\ff_2(u_k,u)
\,\Lambda_2(u)\, t_{12}(u_1) \dots  t_{12}(u_M)\, \Omega 
 \nonumber \\
&&+\, \sum_{k=1}^{M}  Q_k(u;\{u_{j}\})\, t_{12}(u_1) \dots 
t_{12}(u_k\to u)
\dots t_{12}(u_M)\, \Omega \qquad\\
 Q_k(u;\{u_{j}\}) &=& \fg^{-}_{2}(u_k,u) \prod_{j\neq k}^{M} 
\ff_2(u_j,u_k)\,\Lambda_2(u_k).
\label{DBBO}
\een

Demanding $\Phi(\{u\})$ to be an eigenvector of $t(u)$ leads to:
\beq
(-1)^{[1]}P_k(u;\{u\})+ (-1)^{[2]}Q_k(u;\{u\})=0,
\eeq
which corresponds to the cancelling of the so-called `unwanted terms'
carried by the terms $t_{12}(u_1) \dots t_{12}(u_k\to u)\dots t_{12}(u_M)$.
In this way, we get
the Bethe equations 
\beq
 \frac{\Lambda_1(u_k)}{\Lambda_2(u_k)}= 
\prod_{j\neq k}^{M}\frac{\ff_2(u_k,u_j)}{\ff_1(u_j,u_k)}=
 (-1)^{M-1}\prod_{j\neq  k}^{M}\frac{\fa_{2}(u_j,u_k)}{\fa_{1}(u_k,u_j)}\,,
\ k=1,\ldots,M.
\eeq
Remark that the r.h.s. depends only on the structure constants of the 
(super)algebra under consideration, while the l.h.s. encodes the 
representations entering the spin chain.

Then, the eigenvalues of the transfer 
matrix read  
\ben
 t(u)\,\Phi(\{u\}) &=& \Big((-1)^{[1]}t_{11}(u) + 
(-1)^{[2]}t_{22}(u)\Big) \, \Phi(\{u\})  =
\Lambda(u;\{u\})\, \Phi(\{u\}) \\
\Lambda(u;\{u\}) &=& (-1)^{[1]} \Lambda_1(u) \prod_{k=1}^{M} 
\ff_1(u,u_{k})
+(-1)^{[2]} \Lambda_2(u) \prod_{k=1}^{M} \ff_2(u_{k},u)\,.
\een

Note that Bethe equations correspond to the vanishing
of the residue of  $\Lambda(u;\{u\})$. 
This is the tool used in 
analytical Bethe ansatz to obtain Bethe equations, 
see e.g. \cite{byebye,ACDFR2}.


\section{Nested Bethe Ansatz\label{sec:NBA}}

The method, called the Nested Bethe Ansatz (NBA), consists in a
recurrent application of the ABA to express higher 
rank solutions using the lower ones. It has been introduced in 
\cite{KuResh83}.
In this way we can compute the eigenvalues, eigenvectors and Bethe
equations of the $\cA_{\fm|\fn}$ model from the ones of 
$\cA_2$ or $\cA_{1|1}$ model. 

Although we are in a 
(tensor product of) representation(s) of 
$\cA_{\fm|\fn}$, we will loosely keep writing $t_{ij}(u)$ the 
representation of the operators $t_{ij}(u)$, assuming that the 
reader will understand that when $t_{ij}(u)$ applies to the highest weight 
$\Omega$, it is in fact its (matricial) representation that is used.

\subsection{Preliminaries}

As a starter, we decompose the monodromy matrix in the following form (in the auxiliary space
$End(\mathbb{C}^{\fm+\fn}$)):
\beq
T(u) =
 \left ( \begin{array}{cc} 
t_{11}(u) & B^{(1)}(u)  \\ 
C^{(1)}(u)  & T^{(2)}(u) 
\end{array} \right)
\eeq
where $B^{(1)}(u)$ (resp. $C^{(1)}(u)$) is a row (resp. column) vector 
of $\CC^{\fm+\fn-1}$, and $T^{(2)}(u)$ is a matrix of $End(\CC^{\fm+\fn-1})$.

Then, $T^{(2)}(u)$ is itself decomposed in the same way, and more 
generally, for a given $k$ in $\{1,\dots,\fm+\fn-2\}$,
we gather  the generators
$t_{kj}(u)$, (resp. $t_{jk}(u)$) $j=k+1,\ldots,\fn+\fm$,  in a row 
(resp. column) vector of 
$\CC^{\fm+\fn-k}$
and $t_{ij}(u)$, $i,j\geq k$, into a matrix of $End(\CC^{\fm+\fn-k})$:
\begin{eqnarray}
B^{(k)}(u) &=& \sum_{j=k+1}^{\fm+\fn}e^t_{j} \otimes t_{kj}(u)
\mb{and} 
C^{(k)}(u) = \sum_{j=k+1}^{\fm+\fn}e_{j} \otimes t_{jk}(u)
\\
T^{(k+1)} (u) &=& \sum_{i,j=k+1}^{\fm+\fn}\,E_{ij} \otimes t_{ij}(u)
\\[2.1ex]
T^{(k)}(u) &=& \left ( \begin{array}{cc} 
t_{kk}(u) & B^{(k)}(u)  \\ 
C^{(k)}(u)  & T^{(k+1)}(u) 
\end{array} \right)\,.
\label{eq:mono-k}
\end{eqnarray}

We decompose the transfer matrix in the same way:
\ben
t(u) &=& t^{(1)}(u) = (-1)^{[1]}t_{11}(u)+t^{(2)}(u)\,,\nonu
t^{(k)}(u) &=& str\Big(T^{(k)}(u)\Big) = 
(-1)^{[k]}t_{kk}(u)+t^{(k+1)}(u)\,.
\een

At each step of the recursion, the relations  
between $t^{(k)}(u)$, $T^{(k)}(u)$ and $B^{(k)}(u)$ remain similar:
\ben
B^{(k)}_1(u)\,B^{(k)}_2(v) &=& (-1)^{[k]}\frac{\fa_{k+1}(u,v)}{\fa_{k}(u,v)}
B^{(k)}_2(v)\,B^{(k)}_1(u)\,\RR_{12}^{(k+1)}(u,v) 
\label{eq:comBB}
\een
\ben
t_{kk}(u)\,B^{(k)}(v) &=& \ff_{k}(u,v)\,B^{(k)}(v)\,t_{kk}(u) + 
\fg^{+}_{k}(u,v)\,B^{(k)}(u)\,t_{kk}(v) 
\een
\ben
T_1^{(k+1)} (u)\,B^{(k)}_2(v) &=& \ff_{k+1}(v,u)\,B^{(k)}_2 
(v)\,T_1^{(k+1)} (u)\,\RR_{12}^{(k+1)}(u,v) \nonumber\\[1.2ex]
&&+ \fg^{-}_{k+1}(v,u)\,B^{(k)}_2 (u)\,T_1^{(k+1)} (v)\,
 \RR_{12}^{(k+1)}(u,u) \qquad
\label{NCR}\\[1.2ex]
\RR_{12}^{(k)}(u,v) \,T_1^{(k)} (u)\,T_2^{(k)} (v) &=&
T_2^{(k)}(v)\,T_1^{(k)}(u)\,\RR_{12}^{(k)}(u,v)\,.
\een
These relations are proven using the RTT relations (\ref{RTT}) and the 
Yang--Baxter equation (\ref{YBE}).
When $k=\fm+\fn-1$, one recovers the commutation 
relations of $\cA_2$ or $\cA_{1|1}$.

At each step $k=1,\ldots,\fm+\fn-1$ of the nesting, we will introduce a family of spectral 
parameters 
$u^{(k)}_{j}$, $j=1,\ldots,M_{k}$, 
the number $M_{k}$ of these parameters being a free integer. The 
partial unions of these families will be noted as
\begin{equation}
\{u^{(\ell)}\}=\bigcup_{k=1}^{\ell}\,
\{u^{(k)}_{j}\,,\ j=1,\ldots,M_{k}\}
\end{equation}
so that the whole family of spectral parameters is
$\{u\}=\{u^{(\fm+\fn-1)}\}$.

These parameters correspond to the different pseudo-excitations 
above the pseudo-vacuum, and the cardinal of $\{u\}$,  
$M=\sum_{k=1}^{\fm+\fn-1}M_k$, is the total number of these pseudo-excitations. 
Let us stress that, in the same way the pseudo-vacuum is 
\underline{not} the (physical) ground state of the spin chain, these 
pseudo-excitations (above the pseudo-vacuum) are \underline{not} 
physical excitations. However, they do describe states and 
even it is believed/proven (depending on the cases) that they 
describe \textsl{all} the states of the chain.

\subsection{First step of the construction}
{From} the definition of the highest weight, we have
\ben
C^{(1)}(u)\,\Omega = 0
\een
and we can use $B^{(1)}(u)$ as a creation operator. However, since 
$B^{(1)}(u)$ contains only $t_{1j}(u)$ operators, it is clear that we 
need to act on several vectors to describe the whole 
representation with highest weight $\Omega$. The NBA 
spirit is to 
construct these different vectors as Bethe vectors of an
$\cA_{\fm-1|\fn}$ chain that is related to the chain we start with.

More generally, at 
each step $k$ corresponding to the decomposition (\ref{eq:mono-k})
of the monodromy 
matrix, we use (a suitable refinement of) $B^{(k)}(u)$ as a creation operator acting on a 
 set of (to be defined) vectors. These vectors are constructed as 
 Bethe vectors of an $\cA_{\fm-k-1|\fn}$ chain.

At the first step of the recursion, the Bethe vectors have the 
form:  
\ben
\Phi(\{u\}) &=& B^{(1)}_{a_1}(u^{(1)}_1) \dots 
B^{(1)}_{a_{M_1}}(u^{(1)}_{M_1})\,F^{(1)}_{a_1\dots 
a_{M_1}}(\{u\})\,\Omega \\
F^{(1)}_{a_1\dots a_{M_{1}}}(\{u\}) &\in&
 (\CC^{\fm-1|\fn})^{\otimes M_1} \otimes \cA_{\fm-1|\fn}
\een
where $F^{(1)}_{a_1\dots a_{M_{1}}}(\{u\})$ is built  
from operators $t_{ij}(u)$, $2 \leq i \leq j \leq \fm+\fn$ only.
Since $B^{(1)}(u)$ belongs to $\CC^{\fm-1|\fn}\otimes \cA_{\fm|\fn}$, 
we have introduced in the construction $M_{1}$ additional auxiliary spaces 
(labelled $a_{1},\ldots,a_{M_{1}}$) that are also carried by 
$F^{(1)}_{a_1\dots a_{M_{1}}}(\{u\})$. These new auxiliary spaces take 
care of the linear combination one has to do between the different 
generators $t_{1j}(u)$, $j=2,\ldots,\fm+\fn$, that enter into the 
construction.
In the next step of the 
recursion, these new auxiliary spaces 
are re-interpreted as new \textsl{quantum} spaces (i.e. new sites) 
in the fundamental representation of an $\cA_{\fm-1|\fn}$ chain. We 
come back on this point later.

Since $F^{(1)}_{a_1\dots a_{M_{1}}}(\{u\})$ is built up from 
operators $t_{ij}(u)$, 
$2 \leq i \leq j \leq \fm+\fn$, it obeys the relation (proven in a more 
general context in lemma \ref{COM} below)
\ben
t_{11}(u)\,F^{(1)}_{a_1\dots a_{M_1}}(\{u\})\,\Omega &=&
\Lambda_1 (u)\,F^{(1)}_{a_1\dots a_{M_1}}(\{u\})\,\Omega 
\een
so that the action of $t_{11}(u)$ on 
$\Phi(\{u\})$ takes the form:
\ben
t_{11} (u)\, \Phi(\{u\}) &=& \Lambda_1(u)\,\prod_{i=1}^{M_1} 
\ff_1(u,u^{(1)}_i)
\,\Phi(\{u\}) + \sum_{j=1}^{M_1} P_j(u;\{u^{(1)}\})\, \Phi_j(\{u\})
\qquad\label{eq:t11Phi}\\
 P_j(u;\{u^{(1)}\}) &=& 
\Lambda_1 (u^{(1)}_j)\,\fg^{+}_{1}(u,u^{(1)}_j) 
\prod_{i \neq j}^{M_1} \ff_1(u^{(1)}_i,u^{(1)}_j)
\een
where $\Phi_j(\{u\})$ is deduced from $\Phi(\{u\})$ 
by the change $u^{(1)}_j\to\,u$. Expression (\ref{eq:t11Phi}) 
is computed as it has been done in section \ref{sec:ABA}: 
$P_1(u;\{u^{(1)}\})$ is easy to compute; the other terms are obtained 
through a reordering of the operators $B^{(1)}(u^{(1)}_{j})$. For 
details, see lemma \ref{REOD} below which deals with the general case.

It remains to compute the action of $t^{(2)}(u)$ on 
$\Phi(\{u\})$. We do it in two stages. We first commute $t^{(2)}(u)$ 
with the operators $B^{(1)}(u^{(1)}_{j})$:
\ben
&&t^{(2)}(u)\,\Phi(\{u\}) \ =\  \prod_{j=1}^{M_1}
\ff_2(u^{(1)}_j,u)\ B^{(1)}_{a_1}(u^{(1)}_1) \dots 
B^{(1)}_{a_{M_1}}(u^{(1)}_{M_1})\,\wt t^{(2)}(u;\{u^{(1)}\})\,
F^{(1)}_{a_1\dots a_{M_1}}(\{u\})\,\Omega \nonu
&&\qquad+ \sum_{j=1}^{M_1} 
\wh Q_j(u;\{u^{(1)}\})\,B^{(1)}_{a_1}(u^{(1)}_1) \dots 
B^{(1)}_{a_{j}}( u)\dots
B^{(1)}_{a_{M_1}}(u^{(1)}_{M_1})\,\wt t^{(2)}(u^{(1)}_{j};\{u^{(1)}\})\,
F^{(1)}_{a_1\dots a_{M_1}}(\{u\})\,\Omega\nonu
&&\wh Q_j(u;\{u^{(1)}\}) \ =\  \fg^{-}_{2}(u^{(1)}_j,u)\,\prod_{i \neq 
j}^{M_1} \ff_2(u^{(1)}_j,u^{(1)}_i)
\een
where we used the notation
\ben
 \wt t^{(2)}(u;\{u^{(1)}\}) &=& str_{a}\Big(T_{a}^{(2)}(u) 
\prod_{j=1}^{\atopn{\longleftarrow}{M_1}} \RR_{a 
,a_j}^{(2)}(u,u^{(1)}_j)\Big) \,.
\een
Again, calculation is done for $\wh Q_1(u;\{u^{(1)}\})$ and then 
generalized to  $\wh Q_j(u;\{u^{(1)}\})$ using the reordering lemma 
\ref{REOD} and the Yang-Baxter equation.

As already mentionned, the calculation makes appear a new transfer matrix $\wt 
t^{(2)}(u;\{u^{(1)}\})$ corresponding to an $\cA_{\fm-1|\fn}$ chain with $L+M_{1}$ sites, 
the $M_{1}$
additional sites corresponding to fundamental representations of 
$\cA_{\fm-1|\fn}$. This interpretation is supported by the relations
\ben
&&R_{ab}^{(2)}(u,v)\,\wt{T}_{a}^{(2)}(u;\{u^{(1)}\})\,
\wt{T}_{b}^{(2)}(v;\{u^{(1)}\})\,=\,\wt{T}_{b}^{(2)}(v;\{u^{(1)}\})
\,\wt{T}_{a}^{(2)}(u;\{u^{(1)}\})\,R_{ab}^{(2)}(u,v)\qquad \\
&&\wt{T}_{a}^{(2)}(u;\{u^{(1)}\})\,=T_a^{(2)}(u)\,
\prod_{j=1}^{\atopn{\longleftarrow}{M_1}} 
\RR_{aa_j}^{(2)}(u,u^{(1)}_j) \label{T2-tilde}
\een
which ensure that $\wt{T}_{a}^{(2)}(u;\{u^{(1)}\})$ generates 
$\cA_{\fm-1|\fn}$, and that $\wt t^{(2)}(u;\{u^{(1)}\})$ is indeed a transfer 
matrix which obeys 
\beq
[\wt t^{(2)}(u;\{ u^{(1)}\}),\wt t^{(2)}(v;\{ u^{(1)} \})] = 0\,.
\eeq

Then, if we assume that $F^{(1)}_{a_1\dots a_{M_1}}(\{u\})\,\Omega$ 
is an eigenvector of this new transfer matrix:
\ben
\wt t^{(2)}(u;\{u^{(1)}\})\,F^{(1)}_{a_1\dots a_{M_1}}(\{u\})\,\Omega &=& 
\wt\Gamma^{(2)}(u)\,F^{(1)}_{a_1\dots a_{M_1}}(\{u\})\,\Omega\,,
\label{NPB}
\een
we deduce
\ben
t^{(2)}(u)\,\Phi(\{u\}) &=& \wt\Gamma^{(2)}(u)\, 
\prod_{j=1}^{M_1}\ff_2(u^{(1)}_j,u)\,\Phi(\{u\})+ \sum_{j=1}^{M_1} 
Q_j(u;\{u^{(1)}\})\, \Phi_j(\{u\})
\label{eq:t2Phi} \\
Q_j(u;\{u^{(1)}\}) &=& 
\wt\Gamma^{(2)}(u^{(1)}_j)\,\fg^{-}_{2}(u^{(1)}_j,u)\,\prod_{i \neq 
j}^{M_1} \ff_2(u^{(1)}_j,u^{(1)}_i)\,.
\een
Gathering the relations (\ref{eq:t11Phi}) and (\ref{eq:t2Phi}), we get a first 
expression of the action of $t(u)$ on $\Phi (\{u\})$.
When we cancel in this expression the unwanted terms (carried by $\Phi_{j} 
(\{u\})$), we get the first Bethe 
equation and a first expression of the eigenvalue:
\ben
&&(-1)^{[1]}\,\Lambda_1(u^{(1)}_j)\,\fg^{+}_{1}(u,u^{(1)}_j)\,
\prod_{i\neq j}^{M_1}\ff_1(u^{(1)}_j,u^{(1)}_i)
+\,\wt\Gamma^{(2)} (u^{(1)}_j)\,\fg^{-}_{2}(u^{(1)}_j,u)\,
\prod_{i\neq j}^{M_1}\ff_2(u^{(1)}_i,u^{(1)}_j)= 0 
\qquad\quad\\
&&t(u)\,\Phi (\{u\}) = 
\Big((-1)^{[1]}\Lambda_1(u)\prod_{j=1}^{M_1}\ff_1(u,u^{(1)}_j) +
\wt\Gamma^{(2)}(u)\prod_{j=1}^{M_1} \ff_2(u^{(1)}_j,u)\Big)\, \Phi 
(\{u\})\,.
\een
In the above relations, everything is known \textit{but} the 
eigenvalue $\wt\Gamma^{(2)}(u)$, introduced in (\ref{NPB}), and the 
explicit form of $F^{(1)}_{a_1\dots a_{M_1}}(\{u\})$ ensuring that 
(\ref{NPB}) is indeed satisfied.

Thus, at the end of this first recursion step, we have `reduced' the problem of 
computing an eigenvector $\Phi(\{u\})$ for the transfer matrix $t(u)$ 
of an $\cA_{\fm|\fn}$ chain with $L$ sites to the problem of computing 
an eigenvector $\Phi^{(1)}(\{u\})=F^{(1)}_{a_1\dots a_{M_1}}(\{u\})\,\Omega$ for the
transfer matrix $\wt t^{(2)}(u;\{u^{(1)}\})$ 
of an $\cA_{\fm-1|\fn}$ chain with $L+M_{1}$ sites.

\null

To prepare the second step, it remains to single out the highest 
weights corresponding to the fundamental representations carried by 
the new sites. This is done in the following way.  
\ben
\Phi^{(1)}(\{u\}) &=& F^{(1)}_{a_1\dots a_{M_1}}(\{u\})\,\Omega \nonu
\Phi^{(1)}(\{u\}) &=& \wt B^{(2)}_{a^{2}_1}(u^{(2)}_1;\{u^{(1)}\}) \dots 
\wt B^{(2)}_{a^{2}_{M_{2}}}(u^{(2)}_{M_{2}};\{u^{(1)}\})\,F^{(2)}_{a^{2}_1\dots 
a^{2}_{M^{2}}}(\{u\})\,\Omega^{(2)}\\
\Omega^{(2)} &=& (e^{(1)}_1)^{\otimes\,{M_1}}\otimes\Omega\,,
\een
where $e^{(1)}_1=(1,0,\ldots,0)^t \in \CC^{\fm-1|\fn}$ and $F^{(2)}_{a^{2}_1\dots 
a^{2}_{M_{2}}}(\{u\})$ is built on operators $\wt 
t_{ij}(u;\{u^{(1)}\})$, with $j\geq i>2$. The 
operators $\wt B^{(2)}(u;\{u^{(1)}\})$ play the role, for the 
$\cA_{\fm-1|\fn}$ chain of length $L+M_{1}$, of the operators 
$B^{(1)}(u)$ for the $\cA_{\fm|\fn}$ chain of length $L$. Explicitly, 
they are obtained from the following decomposition of the monodromy 
matrix:
\begin{eqnarray}
\wt T^{(2)}(u;\{u^{(1)}\}) &=& \left ( \begin{array}{cc} 
\wt t_{22}(u;\{u^{(1)}\}) & \wt B^{(2)}(u;\{u^{(1)}\})  \\ 
\wt C^{(2)}(u;\{u^{(1)}\})  & T^{(3)}(u;\{u^{(1)}\}) 
\end{array} \right)
\label{eq:monotilde2}
\end{eqnarray}
where $\wt T^{(2)}(u;\{u^{(1)}\})$ has been defined in (\ref{T2-tilde}). 
Note that if we follow the second step up to the end, we will produce, 
as in the first step, a new monodromy matrix
\begin{equation}
\wt T_{a}^{(3)}(u;\{u^{(2)}\})= T_{a}^{(3)}(u;\{u^{(1)}\}) 
\prod_{j=1}^{\atopn{\longleftarrow}{M_2}} 
\RR_{aa_j}^{(3)}(u,u^{(2)}_j)
\end{equation}
corresponding to a new chain based on $\cA_{\fm-2|\fn}$ and of length 
$L+M_{1}+M_{2}$. We want to stress the difference between the 
monodromy matrix $T^{(3)}(u;\{u^{(1)}\})$ appearing at the begining 
of the second step and the monodromy matrix $\wt 
T_{a}^{(3)}(u;\{u^{(2)}\})$ constructed at the end of the same step.

\subsection{General construction at step $k$}

More generally, the step $k$ starts with the problem
\begin{equation}
\wt  t^{(k)}(u;\{u^{(k-1)}\})\,\Phi^{(k-1)}(\{u\})
= \wt\Gamma^{(k)}(u)\,\Phi^{(k-1)}(\{u\})
\label{eq:tkPhik-1}
\end{equation}
where 
\begin{equation}
\wt t^{(k)}(u;\{u^{(k-1)}\}) = str\Big( \wt 
T^{(k)}(u;\{u^{(k-1)}\})\Big)\,.
\end{equation}
 is the transfer matrix of a $\cA_{\fm+1-k|\fn}$ spin chain of length 
$L+\sum_{j=1}^{k-1} M_{j}$ (obtained from the previous step).
We define
\ben
\Phi^{(k-1)}(\{u\}) &=& F^{(k-1)}_{a^{k-1}_1\dots a^{k-1}_{M_k}}(\{u\})\,\Omega^{(k-1)}
\ =\ \BB^{(k)}(\{u^{(k)}\})\,F^{(k)}_{a^{k}_1\dots 
a^{k}_{M_{k}}}(\{u\})\,\Omega^{(k)}\\
\Omega^{(k)} &=& (e^{(k-1)}_1)^{\otimes{M_{k-1}}}\,\otimes \,\Omega^{(k-1)}\,,
\een
with $e^{(k)}_1=(1,0,\ldots,0)^t \in \CC^{\fm-k|\fn}$.
We have introduced
\ben
\BB^{(k)}(\{u^{(k)}\}) &=& \wt B^{(k)}_{a^{k}_1}(u^{(k)}_1;\{u^{(k-1)}\}) \dots 
\wt B^{(k)}_{a^{k}_{M_{k}}}(u^{(k)}_{M_{k}};\{u^{(k-1)}\})
\label{eq:BB}
\een
where the operators are extracted from the monodromy matrix:
\begin{eqnarray}
\wt T^{(k)}(u;\{u^{(k-1)}\}) &=& \left ( \begin{array}{cc} 
\wt t_{kk}(u;\{u^{(k-1)}\}) & \wt B^{(k)}(u;\{u^{(k-1)}\})  \\ 
\wt C^{(k)}(u;\{u^{(k-1)}\})  & T^{(k+1)}(u;\{u^{(k-1)}\}) 
\end{array} \right)\,.
\label{eq:monotildek}
\end{eqnarray}

\begin{rmk}
\label{rmk:auxspace}
In (\ref{eq:BB}), we have indicated only the auxiliary spaces 
$a^k_{j}$, $j=1,\ldots,M_{k}$. In fact, since $\wt T^{(k)}$ is viewed 
as the monodromy matrix of a spin chain of length 
$L+\sum_{j=1}^{k-1}M_{j}$, the other spaces $a^\ell_{j}$, 
$j=1,\ldots,M_{\ell}$, $\ell<k$, are now quantum spaces. Thus, they 
do not appear explicitly in $\wt T^{(k)}$, as the sites of the 
original spin chain, but obviously this monodromy matrix (and its 
components) does depend on all these spaces.
\end{rmk}

We extract from $\wt t^{(k)}(u;\{u^{(k-1)}\})$ the component 
$\wt t_{kk}(u;\{u^{(k-1)}\})$,
\beq
\wt t^{(k)}(u;\{u^{(k-1)}\}) = (-1)^{[k]}\,\wt{t}_{kk} (u;\{u^{(k-1)}\})
+str\Big(\wt{T}^{(k+1)} (u;\{u^{(k)}\} \Big)\,,
\eeq
and compute its action on the vector $\Phi^{(k-1)}(\{u\})$. 

At the first stage, we commute $\wt 
t_{kk}(u;\{u^{(k-1)}\})$ with the operators 
$B^{(k)}(u^{(k)}_{j};\{u^{(k-1)}\})$:
\ben
&&\wt t_{kk}(u;\{u^{(k-1)}\})\, \Phi^{(k-1)}(\{u\}) \ =\ \prod_{j=1}^{M_k} 
\ff_k(u,u^{(k)}_j)\ \BB^{(k)}(\{u^{(k)}\})\ 
\wt t_{kk}(u;\{u^{(k-1)}\})\, \Phi^{(k)}(\{u\}) \nonu
&&\qquad\qquad + \sum_{j=1}^{M_k} \wh P_j(u;\{u^{(k-1)}\})\ 
\BB^{(k)}_j(u;\{u^{(k)}\}) 
\ \wt t_{kk}(u^{(k)}_{j};\{u^{(k-1)}\})\, \Phi^{(k)}(\{u\}) 
\qquad\qquad\label{eq:tkkPhi}
\een
where we have introduced
\ben
\BB_j^{(k)}(u;\{u^{(k)}\}) &=& 
\wt B^{(k)}_{a^{k}_1}(u^{(k)}_1;\{u^{(k-1)}\}) 
\dots B^{(k)}_{a^{k}_{j}}(u^{(k)}_{j}\to u;\{u^{(k-1)}\})\dots
B^{(k)}_{a^{k}_{M_{k}}}(u^{(k)}_{M_{k}};\{u^{(k-1)}\})\nonu
 \wh P_j(u;\{u^{(k-1)}\}) &=& \fg^{+}_{k}(u,u^{(k)}_j) 
\prod_{i \neq j}^{M_k} \ff_k(u^{(k)}_i,u^{(k)}_j)
\een
The calculation is done directly for $\wh P_1$ by collecting the terms 
containing $\wt t_{kk}(u^{(k)}_{1};\{u^{(k-1)}\})$. It is then generalized 
to $\wh P_j$ thanks to 
the following reordering lemma:
\begin{lemma}
For each $k=1,\ldots,\fm+\fn-1$ and $j=1,\ldots,M_{k}$, we have  
\begin{eqnarray}
\BB^{(k)}(\{u^{(k)}\}) &=& \wt B^{(k)}_j(u^{(k)}_j)\,\wt B^{(k)}_1(u^{(k)}_1) \dots 
\wt B^{(k)}_{j-1}(u^{(k)}_{j-1})\,\wt B^{(k)}_{j+1}(u^{(k)}_{j-1}) \dots
\wt B^{(k)}_{M_{k}}(u^{(k)}_{M_{k}})\nonu
&&\times\prod_{i=1}^{\atopn{\longrightarrow}{j-1}}(-1)^{[j]}\,
\frac{\fa_{j+1}(u^{(k)}_i,u^{(k)}_j)}{\fa_{j}(u^{(k)}_i,u^{(k)}_j)}\,
\RR_{ij}^{(k+1)}(u^{(k)}_i,u^{(k)}_j)
\end{eqnarray} 
where the dependence in $\{u^{(k-1)}\}$ has been omitted in $\wt 
B^{(k)}_{p}$.
\label{REOD}
\end{lemma}
\prf:
Direct calculation using the commutation relations (\ref{eq:comBB}).   
\finprf
Since the new $\RR$-matrices appearing in lemma \ref{REOD} commute with 
$\wt t_{kk}(u^{(k)}_{j};\{u^{(k-1)}\})$, one deduces that all $\wh 
P_j$ polynomials have the same form.

In a second stage, we compute the action of $\wt t_{kk}$ on 
$F^{(k)}\,\Omega^{(k)}$:
\begin{lemma}
\label{COM}
For $k=1,2,\ldots,\fm+\fn-1$, the vector $F^{(k)}_{a_1\dots 
a_M}(\{u\})\, \Omega^{(k)}$ obeys the following
relation:
\ben
\wt t_{kk}(u;\{u^{(k-1)}\})\,F^{(k)}_{a_1\dots a_M}(\{u\})\,\Omega^{(k)}=
\wt \Lambda_k (u;\{u^{(k-1)}\}) 
\,F^{(k)}_{a_1\dots a_M}(\{u\})\,\Omega^{(k)} 
\een
where $\wt \Lambda_k (u;\{u^{(k-1)}\})$ is the weight of the representation 
with highest weight vector $\Omega^{(k)}$:
\ben
\wt t_{kk}(u;\{u^{(k-1)}\})\,\Omega^{(k)}=
\wt \Lambda_k (u;\{u^{(k-1)}\}) \,\Omega^{(k)} 
\een
\end{lemma}
\prf:
For $k < i,j,l $, the commutation relations of $\cA_{\fm|\fn}$  
rewrite:
\ben
t_{kk}(u)\,t_{ij}(v) &=& t_{ij}(v)\,t_{kk}(u) +
(-1)^{([k]+[j])([i]+[j])}\,\frac{\fc_{ik}(v,u)}{\fb(v,u)}
\,t_{kj}(v)\,t_{ik}(u)
\nonumber \\
&&-(-1)^{([k]+[i])([k]+[j])}\frac{\fc_{kj}(v,u)}{\fb(v,u)}
\,t_{kj}(u)\,t_{ik}(v)
\\
t_{lk}(u)\,t_{ij}(v) &=& (-1)^{([i]+[j])([l]+[k])}\Big(1+ \delta_{ik}
\frac{\fa_i(v,u)-\fb(v,u)}{\fb(v,u)}\Big)\,
t_{ij}(v)\,t_{lk}(u)\nonu
&&+(1-\delta_{il})\,\frac{\fc_{il}(v,u)}{\fb(v,u)}\,
(-1)^{([i]+[j])([l]+[k])}
t_{lj}(v)\,t_{ik}(u)
\nonumber \\
&&-(-1)^{([i]+[j])([k]+[j])}\,\frac{\fc_{kj}(v,u)}{\fb(v,u)}
\,t_{lj}(u)\,t_{ik}(v)
\een
Since $F^{(k)}$ contains terms of type $t_{ij}(u)$ with
$k < i \leq j $ only, and because of the property 
\ben
\wt t_{ik}(u;\{u^{(k-1)}\})\, \Omega^{(k)} &=& 0\,,\ i>k
\een
we conclude that $\wt t_{kk}(u;\{u\})$ commutes
with $F^{(k)}$. 

The action of $\wt t_{kk}(u;\{u\})$ 
 on $\Omega^{(k)}$ leads to the result.
\finprf

Gathering equation (\ref{eq:tkkPhi}) and lemma \ref{COM}, we get the 
action of $\wt t_{kk}$ on $\Phi^{(k-1)}(\{u\})$:
\ben
\wt t_{kk}(u;\{u^{(k-1)}\})\, \Phi^{(k-1)}(\{u\}) &=& \prod_{j=1}^{M_k} 
\ff_k(u,u^{(k)}_j)\ \wt \Lambda_k (u;\{u^{(k-1)}\})\,
\Phi^{(k-1)}(\{u\})\nonu
&& + \sum_{j=1}^{M_k} P_j(u;\{u^{(k-1)}\})\ 
\BB^{(k)}_j(u;\{u^{(k)}\})\,\Phi^{(k)}(\{u\})
\label{eq:tkkPhik-1}\\
\Phi^{(k)}(\{u\}) &=&
F^{(k)}_{a^{(k)}_1\dots a^{k}_{M^{k}}}(\{u\})\,\Omega^{(k)}
\qquad\qquad\nonu
 P_j(u;\{u^{(k-1)}\}) &=& \wt \Lambda_k (u^{(k)}_j;\{u^{(k-1)}\})\,\fg^{+}_{k}(u,u^{(k)}_j) 
\prod_{i \neq j}^{M_k} \ff_k(u^{(k)}_i,u^{(k)}_j)\,.
\een

It remains to do the same for 
$t^{(k+1)}(u;\{u^{(k-1)}\})=str\Big( T^{(k+1)}(u;\{u^{(k-1)}\})\Big)$. We first 
commute $t^{(k+1)}(u;\{u^{(k-1)}\})$ with $\BB^{(k)}(\{u^{(k)}\})$ 
using relations (\ref{YBE}) and (\ref{NCR}):
\ben
&&t^{(k+1)}(u;\{u^{(k-1)}\})\,\Phi^{(k-1)}(\{u\})\ =\  
\prod_{j=1}^{M_k}\ff_{k+1}(u^{(k)}_j,u)\ \BB^{(k)}(\{u^{(k)}\})\,
\wt t^{(k+1)}(u;\{u^{(k)}\})\,\Phi^{(k)}(\{u\})\quad\nonu
&&\qquad\qquad+ \sum_{j=1}^{M_k} 
\wh Q_j(u;\{u^{(k)}\})\, \BB_{j}^{(k)}(u;\{u^{(k)}\})\,
\wt t^{(k+1)}(u^{(k)}_j;\{u^{(k)}\})\,\Phi^{(k)}(\{u\})
\label{eq:tkPhi} \\
&&\wh Q_j(u;\{u^{(k)}\}) \ =\ \fg^{-}_{k+1}(u^{(k)}_j,u)\,
\prod_{i \neq j}^{M_k} \ff_{k+1}(u^{(k)}_j,u^{(k)}_i)\,.
\een
It makes appear new monodromy and transfer matrices:
\ben
\wt T_{a}^{(k+1)}(u;\{u^{(k)}\}) &=& T_{a}^{(k+1)}(u;\{u^{(k-1)}\}) 
\prod_{j=1}^{\atopn{\longleftarrow}{M_k}} 
\RR_{a,a_j}^{(k+1)}(u,u^{(k)}_j) \\
\wt t^{(k+1)}(u;\{u^{(k)}\}) &=& 
str_{a}\Big(\wt T_{a}^{(k+1)}(u;\{u^{(k)}\})\Big) \,.
\een
The new monodromy matrix also satisfies the RTT relation
\begin{eqnarray*}
&&\!\!\!\!\!\!\!\!
R_{ab}^{(k+1)}(u,v)\,\wt{T}_{a}^{(k+1)}(u;\{u^{(k)}\})\,
\wt{T}_{b}^{(k+1)}(v;\{u^{(k)}\})\,=\,\wt{T}_{b}^{(k+1)}(v;\{u^{(k)}\})
\,\wt{T}_{a}^{(k+1)}(u;\{u^{(k)}\})\,R_{ab}^{(k+1)}(u,v) 
\end{eqnarray*}
so that the problem 
\begin{equation}
\wt t^{(k+1)}(u;\{u^{(k)}\}) \,\Phi^{(k)}(\{u\}) 
= \wt\Gamma^{(k+1)}(u)\,\Phi^{(k)}(\{u\}) 
\label{eq:chain-k+1}
\end{equation}
is integrable, and defines a $\cA_{\fm-k|\fn}$ spin chain, with 
$L+\sum_{j=1}^{k}M_{j}$ sites.

Assuming the form (\ref{eq:chain-k+1}), we get
\ben
t^{(k+1)}(u;\{u^{(k-1)}\})\,\Phi^{(k-1)}(\{u\}) &=& 
\wt\Gamma^{(k+1)}(u)
\,\prod_{j=1}^{M_k}\ff_{k+1}(u^{(1)}_j,u)\ \Phi^{(k-1)}(\{u\})\nonu
&&+ \sum_{j=1}^{M_k} 
Q_j(u;\{u^{(k)}\})\, \,\BB_{j}^{(k)}(u;\{u^{(k)}\})\,\Phi^{(k)}(\{u\})
\\
Q_j(u;\{u^{(k)}\})  &=&  
\wt\Gamma^{(k+1)}(u^{(k)}_j)\,\fg^{-}_{k+1}(u^{(k)}_j,u)\,
\prod_{i \neq j}^{M_k} \ff_{k+1}(u^{(k)}_j,u^{(k)}_i)\,.
\label{eq:tkPhitot}
\een
Gathering (\ref{eq:tkkPhik-1}) and (\ref{eq:tkPhitot}), and comparing them with 
(\ref{eq:tkPhik-1}), we get the $k^{th}$ Bethe equation and an expression for 
$\wt\Gamma^{(k)}(u)$:
\ben
&&\!\!\!\!\!\!\!\!
(-1)^{[k]}\,\wt\Lambda_k(u^{(k)}_j;\{u^{(k)}\})\,\fg^{+}_{k}(u,u^{(k)}_j)\,
\prod_{i\neq j}^{M_k}\ff_k(u^{(k)}_j,u^{(k)}_i) 
+\wt\Gamma^{(k+1)} (u^{(k)}_j)\,\fg^{-}_{k+1}(u^{(k)}_j,u)\,
\prod_{i\neq j}^{M_k} \ff_{k+1}(u^{(k)}_i,u^{(k)}_j)= 0 
\nonu
&&\wt\Gamma^{(k)}(u) = 
(-1)^{[k]}\,\wt\Lambda_k(u;\{u^{(k)}\})\,\prod_{j=1}^{M_k}\ff_k(u,u^{(k)}_j) +
\wt\Gamma^{(k+1)}(u)\,\prod_{j=1}^{M_k} \ff_{k+1}(u^{(k)}_j,u)\,.
\label{eq:BEk}
\een

\subsection{End of the recursion}
To end the recursion, we remark that 
\begin{equation}
\wt\Gamma^{(\fm+\fn)}(u)=(-1)^{[\fm+\fn]}\,\wt\Lambda_{\fm+\fn}(u;\{u^{(\fm+\fn)}\})
\end{equation}
so that $\wt\Gamma$ is expressed in term of $\wt\Lambda$:

\begin{eqnarray}
\wt\Gamma^{(k)}(u) &=& 
(-1)^{[k]}\,\wt\Lambda_{k}(u;\{u^{(k)}\})\,
\prod_{j=1}^{M_{k}}\ff_{k}(u,u^{(k)}_j)
\label{eq:gammak}
\\
&+& \sum_{\ell=k+1}^{\fm+\fn-1}
(-1)^{[\ell]}\,\wt\Lambda_{\ell}(u;\{u^{(\ell)}\})\,
\left(\prod_{j=1}^{M_{\ell}}\ff_{\ell}(u,u^{(\ell)}_j)\right)
\ \left(\prod_{p=k}^{\ell-1}\,
\prod_{j=1}^{M_{p}}\ff_{p+1}(u^{(p)}_j,u)\right)
\nonu
&+& (-1)^{[\fm+\fn]}\,\wt\Lambda_{\fm+\fn}(u;\{u\})\,
\left(\prod_{p=k}^{\fm+\fn-1}\,
\prod_{j=1}^{M_{p}}\ff_{p+1}(u^{(p)}_j,u)\right)\,.
\nonumber
\end{eqnarray}

It remains to compute the values 
$\wt\Lambda_k(u;\{u^{(k)}\})$. It is done in the following lemma:
\begin{lemma}\label{lem:lambdatilde}
The eigenvalue $\wt\Lambda_k(u^{(k)}_j;\{u^{(k)}\})$ of 
$\wt t_{kk}(u;\{u^{(k-2)}\})$ on $\Omega^{(k-1)}$ is given by
\begin{equation}
\wt\Lambda_k(u;\{u^{(k)}\}) = \Lambda_k(u)\,
\prod_{\ell=1}^{k-2}\ \prod_{j=1}^{M_{\ell}} 
\frac{\fb(u,u_{j}^{(\ell)})}{\fa_{\ell+1}(u,u_{j}^{(\ell)})} = \Lambda_k(u)\,
\prod_{\ell=1}^{k-2}\ \prod_{j=1}^{M_{\ell}} 
\frac{1}{\ff_{\ell+1}(u_{j}^{(\ell)},u)}
\qquad k=1,\ldots,\fm+\fn
\nonumber
\end{equation}
where we have used $t_{kk}(u)\,\Omega=\Lambda_k(u)\,\Omega$ for the 
original spin chain.
\end{lemma}
\prf
For $\ell=1,\ldots,\fm+\fn-1$, we compute:
\ben
&&\left(\prod_{j=1}^{\atopn{\longleftarrow}{M_\ell}} R_{aa_j}^{(\ell+1)}
(u,u^{(\ell)}_{j})\right)\, Ê\big(e_{\ell+1}\big)^{\otimes M_{\ell}} \ =\
\left( \prod_{j=1}^{M_{\ell}}
\fa_{\ell+1}(u,u^{(\ell)}_{j})\right)\,E_{\ell+1,\ell+1}\otimes
\big(e_{\ell+1}\big)^{\otimes M_{\ell}}
\nonu
&&\qquad +\left( \prod_{j=1}^{M_{\ell}}
\fb(u,u^{(\ell)}_{j})\right)\,\sum_{s=\ell+2}^{\fm+\fn}E_{ss}\otimes
\big(e_{\ell+1}\big)^{\otimes M_{\ell}}
\nonu
&&\qquad +\sum_{p=1}^{M_\ell}\,\left( \prod_{j=p+1}^{M_{\ell}}
\fa_{\ell+1}(u,u^{(\ell)}_{j}) \right)\,\left( \sum_{s=\ell+2}^{\fm+\fn}
(-1)^{(p-1)[\ell+1]([s]+[\ell+1])}\,
\fc_{\ell+1,s}(u,u^{(\ell)}_{p}) \right)\,
\left( \prod_{j=1}^{p-1}
\fb(u,u^{(\ell)}_{j})\right)
\nonu
&&\qquad\times\ E_{\ell+1,s}\otimes \big(e_{\ell+1}\big)^{\otimes (p-1)}\otimes e_{s} 
\otimes \big(e_{\ell+1}\big)^{\otimes (M_\ell-p)}
\label{eq:Rek}
\een
where the calculation has been done in $\CC^{\fm|\fn}$ with the 
identification $e^{(\ell)}_{1}\equiv e_{\ell+1}$.

In the product of such terms, we want to select the term(s) 
carried by $E_{kk}$ in the auxiliary space (labelled $a$ in 
equation (\ref{eq:Rek})). Since the matrices $E_{ij}$ appearing in 
(\ref{eq:Rek}) are all upper triangular, this implies that each term 
must be carried by a $E_{kk}$ matrix in space $a$. Denoting by 
$E_{kk}^{(a)}$ such matrix, one deduces
\ben
&&str_{a}\left(E_{kk}^{(a)}\ \prod_{\ell=1}^{\atopn{\longrightarrow}{k-1}}\,
\prod_{j=1}^{\atopn{\longleftarrow}{M_\ell}} R_{aa_j}^{(\ell+1)} 
(u,u^{(\ell)}_{j})\right)\, 
\big(e_{k}\big)^{\otimes M_{k-1}}\,\otimes \cdots \otimes\,
\big(e_{3}\big)^{\otimes M_{2}}\,\otimes\,\big(e_{2}\big)^{\otimes M_{1}}\ =\ 
\nonu
&&\  =\  
\left( \prod_{\ell=1}^{k-2}\,\prod_{j=1}^{M_{\ell}} 
\fb(u,u^{(\ell)}_{j})\right)
\prod_{j=1}^{M_{k-1}} \fa_{k}(u,u^{(k-1)}_{j})\ 
\big(e_{k}\big)^{\otimes M_{k-1}}\,\otimes
 \cdots\otimes\, \big(e_{3}\big)^{\otimes M_{2}}\,\otimes\,\big(e_{2}\big)^{\otimes M_{1}}\qquad
\label{eq:Rek-bis}
\een
Remark that we didn't mention the contribution of the original 
$t_{kk}(u)$: in fact, since $\Omega$ is a highest weight, the 
monodromy matrix $T(u)$ is also upper triangular, so that we need 
also to select only $E^{(a)}_{kk}$ for this term. As a consequence, 
the product of $R$-matrices on its own must be carried by  $E^{(a)}_{kk}$.

Finally, from (\ref{eq:Rek-bis}) and the normalisation (\ref{Runit}), we get the 
result.
\finprf
{From} the expression given in lemma \ref{lem:lambdatilde}, one deduces that:
\begin{eqnarray}
\wt\Gamma^{(k)}(u;\{u\}) &=& \left(\prod_{p=1}^{k-1}\,
\prod_{j=1}^{M_{p}}\frac{1}{\ff_{p+1}(u^{(p)}_j,u)}\right)\,
\left\{
(-1)^{[k]}\,\Lambda_{k}(u)\,
\left(\prod_{j=1}^{M_{k}} \ff_{k}(u,u^{(k)}_j)\right)\,
\left(\prod_{j=1}^{M_{k-1}} \ff_{k}(u^{(k-1)}_j,u)\right)
\right.\nonu
&&+ \sum_{\ell=k+1}^{\fm+\fn-1}
(-1)^{[\ell]}\,\Lambda_{\ell}(u)\,
\left(\prod_{j=1}^{M_{\ell}}\ff_{\ell}(u,u^{(\ell)}_j)\right)
\,\left(\prod_{j=1}^{M_{\ell-1}}\ff_{\ell}(u^{(\ell-1)}_j,u)\right)
\nonu
&&+ \left.(-1)^{[\fm+\fn]}\,\Lambda_{\fm+\fn}(u)\,
\left(\prod_{j=1}^{M_{\fm+\fn-1}}
\ff_{\fm+\fn}(u^{(\fm+\fn-1)}_j,u)\right)\right\}\,.
\label{eq:gammak-tot}
\end{eqnarray}

Let us note that since $\fb(u,u)=0$, equation 
(\ref{eq:gammak-tot}) implies that:
\begin{eqnarray}
&&\wt\Gamma_k(u^{(\ell)}_{j};\{u\}) \ =\ 0
\mb{for} j=1,\ldots,M_{\ell}\,;\ \ell=1,\ldots,k-2 \\
&&\wt\Gamma_k(u^{(k-1)}_{i};\{u\}) \ =\ 
(-1)^{[k]}\,\Lambda_{k}(u)\,
\left(\prod_{j=1}^{M_{k}} \ff_{k}(u^{(k-1)}_{i},u^{(k)}_j)\right)\,
\left(\prod_{p=1}^{k-2}\,
\prod_{j=1}^{M_{p}}\frac{1}{\ff_{p+1}(u^{(p)}_j,u^{(k-1)}_{i})}\right)
\nonu
&& \mb{for}i=1,\ldots,M_{k-1}\,.
\end{eqnarray}

\subsection{Final form of Bethe vectors, eigenvalues and equations}
Using these expressions and the value of 
$\wt\Lambda_{k}(u;\{u^{(k)}\})$ given in lemma \ref{lem:lambdatilde}, 
one can recast the Bethe equation (\ref{eq:BEk}) in its final form:
\ben
\frac{\Lambda_{k+1} (u^{(k)}_j) }{\Lambda_k (u^{(k)}_j) } &=&
(-1)^{M_k}\prod_{i=1}^{M_{k-1}}    
\frac{\fa_k(u_j^{(k)},u_i^{(k-1)})}{\fb (u_j^{(k)},u_i^{(k-1)})}\ 
\prod_{i \neq j}^{M_k} 
\frac{\fa_k(u_i^{(k)},u_j^{(k)})}{\fa_{k+1}(u_j^{(k)},u_i^{(k)})}\
\prod_{i=1}^{M_{k+1}}   
\frac{\fb(u_i^{(k+1)},u_j^{(k)})}{\fa_{k+1}(u_i^{(k+1)},u_j^{(k)})}
\nonu
&& j=1,\ldots,M_{k}\,,\quad k=1,\ldots,\fm+\fn-1
\label{BE}
\een
with the convention $M_{0}=M_{\fm+\fn}=0$.
Remark that, in the distinguished gradation, one can simplify 
these equations, see section \ref{sect:ex-super}.

The eigenvalue of the transfer matrix is obtained from 
(\ref{eq:gammak-tot}), remarking that $\Lambda(u)=\wt\Gamma^{(1)}(u)$:
\ben
\Lambda(u)&=&\sum_{k=1}^{\fm+\fn} (-1)^{[k]}\Lambda_{k}(u) 
\prod_{j=1}^{M_{k-1}} \ff_k(u_j^{(k-1)},u)\ \prod_{j=1}^{M_{k}} 
\ff_k(u,u_j^{(k)}) \,.\label{eq:Lambdafin}
\een
Again, due to the distinguished gradation, one can
 simplify the expression of $\Lambda(u)$.

 The number of  parameter families is $\fm+\fn-1$. The Bethe equations
(\ref{BE}) ensure that $\Lambda(u)$ is analytical, in accordance with
the analytical Bethe ansatz.

The Bethe vectors take the form
\ben
\Phi(\{u\}) &=& B^{(1)}_{a_1}(u^{(1)}_1) \cdots 
B^{(1)}_{a_{M_1}}(u^{(1)}_{M_1})\,F^{(1)}_{a_1\dots 
a_{M_1}}(\{u\})\,\Omega 
\label{eq:Phi51}\\[1.2ex]
&=& B^{(1)}_{a^1_1}(u^{(1)}_1)\cdots B^{(1)}_{a^1_{M_1}}(u^{(1)}_{M_1})
\,\wt B^{(2)}_{a^2_2}(u^{(2)}_1)\cdots\wt B^{(2)}_{a^2_{M_2}}(u^{(2)}_{M_2})
\cdots\wt 
B^{(\fn+\fm-1)}_{a^{\fn+\fm-1}_{M}}(u^{(\fn+\fm-1)}_{M})
\,\Omega^{(\fn+\fm-1)}\,.\nonumber
\een
We remind the notation $M=\sum_{j=1}^{\fn+\fm-1}M_{j}$, 
$\Omega^{(k)}= \big(e^{(k-1)}_{1}\big)^{\otimes M_{k-1}}\,\Omega^{(k-1)}$, 
$\Omega^{(1)}=\Omega$ and the auxiliary 
spaces are indicated according to remark \ref{rmk:auxspace}. 

\subsection{Bethe equation in the distinguished gradation\label{sec:dist-gr}}
For this grade, the properties
\begin{eqnarray}
\fa_{k}(u,v) &=& \fa_{1}(u,v)\equiv \fa(u,v) \mb{for} k\leq\fm \mb{ and }
\fa_{k}(u,v) = -\fa(v,u) \mb{for} k>\fm \qquad
\\
\ff_{k}(u,v) &=& \ff_{1}(u,v)\equiv \ff(u,v) \mb{for} k\leq\fm \mb{ and }
\ff_{k}(u,v) = \ff(v,u) \mb{for} k>\fm 
\end{eqnarray}
allow to simplify the Bethe equations to the following form
\ben
\frac{\Lambda_{2} (u^{(1)}_j) }{\Lambda_1 (u^{(1)}_j) } &=&
-\prod_{i \neq j}^{M_1} 
\frac{\ff(u_j^{(1)},u_i^{(1)})}{\ff(u_i^{(1)},u_j^{(1)})}\
\prod_{i=1}^{M_{2}}\Big(\ff(u_j^{(1)},u_i^{(2)})\Big)^{-1}
\qquad
j = 1,\ldots,M_{1}\,,\\[1.7ex]
\frac{\Lambda_{k+1} (u^{(k)}_j) }{\Lambda_k (u^{(k)}_j) } &=&
-\prod_{i=1}^{M_{k-1}} \ff(u_i^{(k-1)},u_j^{(k)})\ 
\prod_{i \neq j}^{M_k} 
\frac{\ff(u_j^{(k)},u_i^{(k)})}{\ff(u_i^{(k)},u_j^{(k)})}\
\prod_{i=1}^{M_{k+1}}   
\Big(\ff(u_j^{(k)},u_i^{(k+1)})\Big)^{-1}
\nonu
j &=& 1,\ldots,M_{k}\,,\quad k=2,\ldots,\fm-1\\[1.7ex]
\frac{\Lambda_{\fm+1} (u^{(\fm)}_j) }{\Lambda_{\fm} (u^{(\fm)}_j) } &=&
-\prod_{i=1}^{M_{\fm-1}}\ff(u_i^{(\fm-1)},u_j^{(\fm)})\ 
\prod_{i=1}^{M_{\fm+1}}   
\Big(\ff(u_i^{(\fm+1)},u_j^{(\fm)})\Big)^{-1}
\qquad j = 1,\ldots,M_{\fm}\qquad\quad\\[1.7ex]
\frac{\Lambda_{k+1} (u^{(k)}_j) }{\Lambda_k (u^{(k)}_j) } &=&
-\prod_{i=1}^{M_{k-1}}    
\ff(u_j^{(k)},u_i^{(k-1)})\ 
\prod_{i \neq j}^{M_k} 
\frac{\ff(u_i^{(k)},u_j^{(k)})}{\ff(u_j^{(k)},u_i^{(k)})}\
\prod_{i=1}^{M_{k+1}}   
\Big(\ff(u_i^{(k+1)},u_j^{(k)})\Big)^{-1}
\nonu
j &=& 1,\ldots,M_{k}\,,\quad k=\fm+1,\ldots,\fm+\fn-2\\[1.7ex]
\frac{\Lambda_{\fm+\fn} (u^{(\fm+\fn-1)}_j) }{\Lambda_{\fm+\fn-1} (u^{(\fm+\fn-1)}_j) } &=&
-\prod_{i=1}^{M_{\fm+\fn-2}}\ff(u_j^{(\fm+\fn-1)},u_i^{(\fm+\fn-2)})\ 
\prod_{i \neq j}^{M_{\fm+\fn-1}} 
\frac{\ff(u_i^{(\fm+\fn-1)},u_j^{(\fm+\fn-1)})}
{\ff(u_j^{(\fm+\fn-1)},u_i^{(\fm+\fn-1)})}\nonu
j &=& 1,\ldots,M_{\fm+\fn-1}\,.
\een
The Bethe equations depend on the highest 
weights $\Lambda_{j}(u)$ and on a sole function: 
\begin{equation}
\ff(u,v) = \frac{\fa(v,u)}{\fb(v,u)}=
\begin{cases}\displaystyle
\frac{u-v+\hbar}{u-v} &\mb{for super-Yangians} \\[1.7ex]
\displaystyle
\frac{q^{-1}\,u^2-q\,v^2}{u^2-v^2}
&\mb{for deformed superalgebras} \end{cases} 
\label{eq:deff}
\end{equation}
It is also true for the transfer matrix eigenvalue:
\ben
\Lambda(u) &=& \Lambda_{1}(u)\ \prod_{j=1}^{M_{1}}\ff(u,u_j^{(1)})
+\sum_{k=2}^{\fm} \Lambda_{k}(u) 
\prod_{j=1}^{M_{k-1}} \ff(u_j^{(k-1)},u)\ \prod_{j=1}^{M_{k}} 
\ff(u,u_j^{(k)})\nonu
&& -\sum_{k=\fm+1}^{\fm+\fn-1} \Lambda_{k}(u) 
\prod_{j=1}^{M_{k-1}} \ff(u,u_j^{(k-1)})\ \prod_{j=1}^{M_{k}} 
\ff(u_j^{(k)},u) - \Lambda_{\fm+\fn}(u) 
\prod_{j=1}^{M_{{\fm+\fn}-1}} \ff(u,u_j^{({\fm+\fn}-1)}) 
\,.\qquad
\label{eq:Lambda-dist}
\een
\subsection{Cartan eigenvalues of Bethe vectors}
It was shown in \cite{byebye,RS,ACDFR2} that the transfer matrix $t(u)$ 
commutes with the Cartan subalgebra of $\cB_{\fm|\fn}$. Hence, Bethe 
vectors are also eigenvectors of the Cartan generators. We give 
hereafter their eigenvalues. Let us  remark that when $\cA_{\fm|\fn}=\cY(\fm|\fn)$
(or $\cY(\fn)$) the symmetry algebra extends to the whole $\cB_{\fm|\fn}$ 
algebra. We remind that we note 
$\lambda^{\langle k\rangle}=(\lambda_{1}^{\langle k\rangle},\ldots,
\lambda_{\fm+\fn}^{\langle k\rangle})$ the 
$\cB_{\fm|\fn}$ highest weight at site $k$.

For  super Yangian $\cY(\fm|\fn)$, the $\cB_{\fm|\fn}$ Cartan generators  
have the form:
\begin{eqnarray}
t_{jj}^{(1)} &=& -(-1)^{[j]}\,\hbar\,\sum_{k=1}^L \II ^{\otimes ^{k-1}} \otimes \cE_{jj} \otimes 
\II ^{\otimes ^{L-k-1}}\\
t_{jj}^{(1)}\,\Phi(\{u\}) &=& -(-1)^{[j]}\,\hbar\,
\Big(M_{j-1}-M_j+\sum_{k=1}^L\lambda^{\langle k\rangle}_j\Big)\, 
\Phi(\{u\})\,.
\end{eqnarray}

For the  super quantum affine algebra $\hat\cU_q(\fm|\fn)$, the $\cB_{\fm|\fn}$ Cartan 
generators are given by:
\begin{eqnarray}
l_{jj}^\pm &=&(-1)^{[j]\,L}\,(q^{\pm H_{j}})^{\otimes^{L}} 
\equiv (-1)^{[j]\,L}\,q^{\pm h_{j}}
\\
q^{h_{j}}\, \Phi(\{u\})&=&\Big(\prod_{\ell=1}^{L}\eta_{\ell}\Big)\,q^{(1-2[j])(M_j-M_{j-1})+ 
\sum_{k=1}^L\lambda^{\langle k\rangle}_j}\,\Phi(\{u\})\,.
\end{eqnarray}

\section{Form of the Bethe vectors\label{sec:betheV}}

In this section, we make contact with the expressions obtained 
in \cite{MTV,TaVa} for Bethe 
vectors of $\cY(\fn)$ and $\wh\cU_{q}(\fn)$ chains. Note that the 
construction there is quite the same, but the proof is rather 
different. We have chosen to stick to the original NBA formalism with 
a constructive approach for the Bethe vectors. In this section, we show how to 
reproduce  some of the results 
given in \cite{MTV,TaVa}, such as the recursion formula for Bethe 
vector and the `trace form' which is the central result of these papers. 
We also generalise them to the case of superalgebras.

\subsection{Recursion formula for Bethe vectors}

{From} expression (\ref{eq:Phi51}), we can extract a recurrent form 
for the Bethe vectors:
\ben
\Phi^{\fn+\fm}_M(\{u\}) &=& B^{(1)}_{a^1_1}(u^{(1)}_1) \cdots 
B^{(1)}_{a^1_{M_1}}(u^{(1)}_{M_1})\,\wh\Psi_{\{u^{(1)}\}}
\Big(\Phi^{{\fn+\fm}-1}_{M-M_{1}}(\{u^{(>1)}\})\Big)
\label{eq:recurPhi}\\
\wh\Psi_{\{u^{(1)}\}} &=& 
v^{(2)}\,\circ \,
(\psi \otimes \pi_{u^{(1)}_{M_1}} \otimes \dots \otimes 
\pi_{u^{(1)}_1}) \circ \Delta^{(M_1)}
\een
where $\pi_{a}$ is the fundamental representation evaluation 
homomorphism normalized as: 
\ben
\pi_a: \begin{array}{lcl}   
\cA_{\fm|\fn}\otimes End(\CC^{\fm|\fn}) & \to &  
End(\CC^{\fm|\fn})\otimes End(\CC^{\fm|\fn})\\
T(u) & \mapsto & \displaystyle \,\RR(u,a)
\end{array}
\een
$v^{(k)}$ is the application of the highest weight vector from the right, 
\begin{equation}
v^{(k)}(X\,\Omega)=X\,(e_{1}^{(k)})^{\otimes M_{k-1}}\otimes\,\Omega\,,
\end{equation}
and $\psi$ is the embedding of $\cA_{\fm-1|\fn}$ in $\cA_{\fm|\fn}$ given 
by
\ben
\psi: \begin{array}{lcl}  
\cA_{\fm-1|\fn} & \to & \cA_{\fm|\fn} \\[1.2ex]
t_{ij}(u) & \mapsto &   t_{i+1,j+1}(u)\\
\end{array}
\een
If we denote by $[.]_{\fm|\fn}$ the grading used in the 
$\cA_{\fm|\fn}$ superalgebra, the embbeding $\psi$ corresponds to the 
identification $[j]_{\fm-1|\fn}=[j+1]_{\fm|\fn}$.

Expression (\ref{eq:recurPhi}) has been given in \cite{MTV,TaVa} in 
the case of $\cY(\fn)$ and $\wh\cU_{q}(\fn)$ chains. It is also valid in the case 
of $\cY(\fm|\fn)$ and $\wh\cU_{q}(\fm|\fn)$ superalgebras.

\subsection{Supertrace formula for Bethe vectors\label{sect:supertrace}}

We can also write the Bethe vector into a supertrace formula and prove the equivalence with the 
recurrence relation discussed above.

\ben
\Phi^{\fn+\fm}_M(\{u\}) &=& (-1)^{A_{1}}
str_{1\ldots M} \,\left(T_1(u_1^{(1)})\dots 
T_M (u_{M_{{\fn+\fm}-1}}^{(\fn+\fm-1)})\,
\RR_{1\dots M}(\{u\})\,\right. \nonu
&&\quad\qquad\qquad\qquad\left.\times\,
 E_{{\fn+\fm},{\fn+\fm}-1}^{\vec{\otimes}M_{\fn+\fm-1}} 
\otimes \dots \otimes  
E_{21}^{\vec{\otimes}{M_1}}\right) \,\Omega 
\label{eq:Phi-str}\\
\RR_{1\dots M}(\{u\}) &=&\prod_{j<k}
\prod_{l=1}^{\atopn{\longrightarrow}{M_k}} 
\prod_{i=1}^{\atopn{\longleftarrow}{M_j}}
\RR_{a_l^k a_i^j}(u_l^{(k)},u_i^{(j)})
\frac{\fa_{1}(u_l^{(k)},u_i^{(j)})}{\fa_{k}(u_l^{(k)},u_i^{(j)})}
\label{eq:bigR} \\
A_{k} &=& \sum_{i=k}^{{\fn+\fm}-2}\frac{M_i(M_i+1)}{2}[i]
\label{eq:defAk}
\een 
We note $1, \dots, M $ the ordered sequence of auxiliary spaces
$a_1^1, \dots, a_{M_1}^1, \, a_1^2, \dots,
a_{M_{\fm+\fn-1}}^{\fm+\fn-1}$.
When $[i]=0$, we recognize the expression given in \cite{MTV,TaVa} for the 
Yangian $\cY(\fn)$ and for the quantum group 
$\wh\cU_{q}(\fn)$. The above expression is also valid in the case 
of $\cY(\fm|\fn)$ and $\wh\cU_{q}(\fm|\fn)$ superalgebras.

Equivalence is proven along the following lines.
Starting from expression (\ref{eq:Phi-str}), we can extract the 
$M_1$ auxiliary spaces
corresponding to the first step of the nested Bethe ansatz  :
\begin{eqnarray*}
\Phi^{\fn+\fm}_M(\{u\}) &=& 
(-1)^{\frac{M_1(M_1+1)}{2}[1]}\, str_{1\ldots M_1} 
\Big[T_1(u_1^{(1)}) \dots T_{M_1}(u_{M_1}^{(1)})\, (-1)^{A_{2}}\,
str_{M_1+1 \ldots M}\,\Big(\qquad
\nonumber \\
&&T_{M_1+1}(u_1^{(2)}) \dots 
T_M (u_{M_{{\fn+\fm}-1}}^{(\fn+\fm-1)})\,
\RR_{1\dots M}(\{u\})\,  
E_{{\fn+\fm},{\fn+\fm}-1}^{\vec\otimes{M_{\fn+\fm-1}}} 
\otimes \dots \otimes E_{32}^{\vec\otimes{M_2}} \Big) \,
\nonu
&&\otimes \, E_{21}^{\vec\otimes{M_1}} \Big] \,\otimes \Omega 
\qquad\qquad\qquad
\end{eqnarray*}
Using the isomorphism
$End(\CC^{\fm+\fn}) \sim \CC^{\fm+\fn}\otimes \CC^{\fm+\fn}$,
one can rewrite, for any $A(v)$, the supertrace 
with an $E_{21}$ matrix as
\begin{eqnarray}
str\Big(T(u)\,A(v)\,E_{21}\Big) 
&=& \sum_{j=1}^{\fm+\fn} \big( e_{1}^t\otimes e_{j}^t\otimes t_{1j}(u)\big)
\,A(v)\, \big(e_{1}\otimes e_{2}\otimes 1\big) 
\label{eq:str-form}\\
&=& (-1)^{[1]+[1]\,[A]}
\sum_{j=1}^{\fm+\fn} \big(e_{j}^t\otimes t_{1j}(u)\big)\,A(v)
\, \big(e_{2}\otimes 1\big)\,.
\end{eqnarray}

Using formula (\ref{eq:str-form}) for the auxiliary spaces 
$1,\ldots,M_{1}$, and remarking that the case 
$j_a=1$ for $a =1, \dots, M_1$ does not contribute, we obtain:
\ben
\Phi^{\fn+\fm}_M(\{u\}) &=& 
B^{(1)}_{a^1_1}(u^{(1)}_1)\cdots 
B^{(1)}_{a^1_{M_1}}(u^{(1)}_{M_1})
\,(-1)^{A_{2}}\,str_{M_1+1 \ldots M}\,\Big(T_{M_1+1}(u_1^{(2)}) \dots 
T_M (u_{M_{{\fn+\fm}-1}}^{(\fn+\fm-1)})\,
\nonumber \\
&&\times \RR_{1\dots M}(\{u\})\,  
E_{{\fn+\fm},{\fn+\fm}-1}^{\vec\otimes {M_{{\fn+\fm}-1}}} 
\otimes \dots \otimes E_{32}^{\vec\otimes {M_2}} \Big) \,
\Omega^{(2)}
\een

To end the proof, we make the following mappings:
\begin{eqnarray}
\cA_{\fm|\fn} &\to& \cA_{\fm-1|\fn}
\\
{[i]}_{\fm|\fn} &\to& [i-1]_{\fm-1|\fn}
\\
\RR_{a^i_ja_k^1}(u_i^{(j)},u_k^{(1)}) &\to& 
\pi_{u^{(1)}_{k}} (T_{a^i_j}(u_i^{(j)}))\\
E_{j+1,j}\in\CC^{\fm|\fn} &\to& E_{j,j-1}\in\CC^{\fm-1|\fn} 
\end{eqnarray}
they allow to recover the definition of $\wh\Psi_{\{u^{(1)}\}}$
and the form (\ref{eq:recurPhi}).

\subsection{Orthogonality relation for Bethe vectors}

In this part we prove the condition for the orthogonality of the 
on-shell Bethe vectors (i.e. when Bethe equations are satisfied).

Let $\cF$ be the space of all Bethe vectors.
 We introduce the Shapovalov form \cite{MTV,KAC}:
\begin{equation}
\langle.\,,\,.\rangle\ :\ \cF \otimes \cF \to \CC
\end{equation}
  which obeys the following properties:
\ben
\langle \, \Omega \,,\, \Omega \, \rangle &=&  1\,, \mb{where $\Omega$ 
is the  highest weight vector of $gl(\fm|\fn)$,}
\\
\langle \,  t_{ij}(u)\, \omega_{1} \,,\, \omega_{2} \rangle &=& \langle \omega_{1} \,,\, 
t_{ji}(u)\, \omega_{2} \rangle \qquad \forall\ 
\omega_{1},\omega_{2}\in\cF.
\label{SF}
\een

\begin{proposition}
$\langle \Phi(\{u\}) \,,\,\Phi(\{v\}) \rangle$ is different from zero 
if and only if  $\{u^{(k)}\}=\{v^{(k)}\}$, $\forall 
k=1,\ldots,\fm+\fn-1$, the sets being not ordered.
\end{proposition}
\prf
{From} the  eigenvalues of $t(u)$ computed within the NBA method we have
\ben
&& t(w)\,\Phi(\{v\})=\Lambda(w;\{v\})\,\Phi(\{v\}) \\[1.2ex]
&& \langle t(w) \Phi(\{u\}) \,,\,\Phi(\{v\}) \rangle = \langle \Phi(\{u\}) 
\,,\,
  t(w) \Phi(\{v\}) \rangle \\[1.2ex]
&& \Lambda(w;\{u\})\, \langle \Phi(\{u\}) \,,\,\Phi(\{v\}) \rangle = 
\Lambda(w;\{v\})\, \langle \Phi(\{u\}) \,,\,\Phi(\{v\}) \rangle 
\een
where $\{u\}$ and $\{v\}$ refer to two different sets of parameters for 
the Bethe vector.
Thus, in order to get $ \langle \Phi(\{u\}) \,,\,\Phi(\{v\}) \rangle $ different 
from zero, we must have:
\ben
  \Lambda(w;\{u\})= \Lambda(w;\{v\})\,.
\een
Since this equality must be satisfied for all values of $w$, and looking at (\ref{BE}), 
we conclude that all the families of Bethe 
roots must be the same up to a permutation in each family $M_i$: 
$\{u^{(k)}_i\,,\ i=1,\ldots M_{k}\}=\{v^{(k)}_j\,,\ j=1,\ldots M_{k}\}$ for all $k$. 
\finprf

\subsection{Examples of Bethe vectors}
Using the definition of the Bethe vector (\ref{eq:bigR}), it is easy to compute 
their explicit form in some specific cases. We illustrate it below, 
but a general expression in term
of the generators $t_{ij}(v)$ is still lacking.

\paragraph{Bethe vectors of $\cA_{\fm|\fn}$ with $\fn+\fm = 2$ and $M_1=M$.}
We reproduce here the well-known case obtained with algebraic Bethe 
ansatz. 
\ben
\Phi^{2}_M(\{u^{(1)}\}) = (-1)^{M [2]}\,t_{12}(u^{(1)}_1) \cdots 
t_{12}(u^{(1)}_M)\, \Omega
\label{eq:Bethegl2}
\een
Note that this expression is also valid when $\fn+\fm>2$, setting 
$M_{1}=M$ and $M_{k}=0$, $k>1$.

\paragraph{Bethe vectors of $\cA_{\fm|\fn}$ with $\fn+\fm = 3$, $M_1=M$ 
and $M_2=1$.}
This case is a generalization of the case $M_{1}=M_{2}=1$ done for 
$\cY(gl_{\fn})$ and $\cU_{q}(gl_{\fn})$ in \cite{MTV,TaVa}.

\begin{eqnarray}
&&\Phi^{3}_{M+1}(u^{(1)}_1,\dots,u^{(1)}_M,u^{(2)}_1) \ =\ 
\label{eq:Bethegl21}\\
&&
(-1)^{\frac{(M+1)M}{2}[1]+M[2]} \,
\left(\prod_{i=1}^M
\frac{\fb(u^{(2)}_1,u_i^{(1)})}{\fa_2(u^{(2)}_1,u_i^{(1)})} 
\right)\Big\{
(-1)^{[3]}\,
t_{12}(u^{(1)}_1) \cdots t_{12}(u^{(1)}_M)\,t_{23}(u_1^{(2)}) \nonumber \\
&&\qquad
+(-1)^{[2]}\,\sum_{i=1}^M
\frac{\fc_{23}(u_1^{(2)},u^{(1)}_i)}{\fb(u^{(2)}_1,u_i^{(1)})}
\,\prod_{k=i+1}^{M}\frac{\fa_2(u^{(2)}_1,u_i^{(1)})}{\fb(u^{(2)}_1,u_i^{(1)})}
\nonu
&&\qquad\qquad\times 
t_{12}(u^{(1)}_1) \cdots t_{12}(u^{(1)}_{i-1})\,
t_{13}(u_i^{(1)})\,t_{12}(u^{(1)}_{i+1}) \cdots t_{12}(u^{(1)}_M)\,
t_{22}(u_1^{(2)})\Big\}\,\Omega 
\nonumber
\end{eqnarray}
Again, this expression is also valid when $\fn+\fm>3$, setting 
$M_{k}=0$, $k>2$.

\paragraph{Bethe vectors of $\cA_{\fm|\fn}$ with $\fn+\fm = 4$, 
$M_1=M$ and $M_2=M_3=1$.}
This case is a generalization of the case $M_{1}=M_{2}=M_3=1$ done for 
$\cY(gl_{\fn})$ and $\cU_{q}(gl_{\fn})$ in \cite{MTV,TaVa}.
\ben
&&\Phi^{4}_{M+2}(u^{(1)}_1,\dots,u^{(1)}_M,u^{(2)}_1,u^{(3)}_1) = 
(-1)^{A_1+[2] M_1+[4]} 
\left(\prod_{j=1}^{M_1} 
\frac{\fb(u_1^{(3)},u_j^{(1)})}{\fa_{3}(u_1^{(3)},u_j^{(1)})}\right)\
\left(\prod_{j=1}^{M_1}
\frac{\fb(u_1^{(2)},u_j^{(1)})}{\fa_{2}(u_1^{(2)},u_j^{(1)})}\right)\
\nonu
&&
\times\frac{\fb(u_1^{(3)},u_1^{(2)})}{\fa_{3}(u_1^{(3)},u_1^{(2)})} 
 \Big\{(-1)^{[3]}
\ t_{12}(u_1^{(1)}) \dots t_{12}(u_M^{(1)}) t_{23}(u_1^{(2)}) 
t_{34}(u_1^{(3)})
\, \Omega
\nonumber \\
&&\qquad+(-1)^{[4]}
\frac{\fc_{34}(u_1^{(3)},u_1^{(2)})}{\fb(u_1^{(3)},u_1^{(2)})}
\ t_{12}(u_1^{(1)}) \dots t_{12}(u_M^{(1)})t_{24}(u_j^{(2)}) 
t_{33}(u_1^{(3)})\, \Omega
\nonumber \\
&&\qquad
+\sum_{k=1}^{M_1} (-1)^{[2]}
\frac{\fc_{23}(u_1^{(2)},u_i^{(1)})}{\fb(u_1^{(2)},u_i^{(1)})}
\left(\prod_{j=k+1}^{M_1}
\frac{\fa_{2}(u_1^{(2)},u_k^{(1)})}{\fb(u_1^{(2)},u_k^{(1)})} \right)
\nonumber \\
&&\qquad
\qquad\qquad \times t_{12}(u_1^{(1)}) \dots  t_{13}(u_k^{(1)}) 
\dots t_{12}(u_M^{(1)})
t_{22}(u_1^{(2)})t_{34}(u_1^{(3)})
\, \Omega
\nonumber \\
&&\left.\qquad
+\sum_{k=1}^{M_1}(-1)^{[2]} 
\frac{\fc_{24}(u_1^{(2)},u_k^{(1)})}{\fb(u_1^{(2)},u_k^{(1)})}
\frac{\fc_{34}(u_1^{(3)},u_1^{(2)})}{\fb(u_1^{(3)},u_1^{(2)})}
\left(\prod_{j=k+1}^{M_1}
\frac{\fa_{2}(u_1^{(2)},u_j^{(1)})}{\fb(u_1^{(2)},u_j^{(1)})}\right) 
\right.
\nonumber \\
&&\left.\qquad
\qquad \qquad\times   t_{12}(u_1^{(1)}) \dots t_{14}(u_k^{(1)}) 
\dots t_{12}(u_M^{(1)}) t_{22}(u_1^{(2)})t_{33}(u_1^{(3)})\, \Omega
\right\}
\label{eq:Bethegl22}
\een
where $A_{1}$ is defined in (\ref{eq:defAk}).

\section{Application to AdS/CFT correspondence\label{sect:ex-super}}
To illustrate the technics, we present some Bethe vectors in the case 
of $\cA_{2|1}$, $\cA_{2|2}$ and $\cA_{4|4}$. These superalgebras, 
when they are undeformed, appeared recently in the AdS/CFT 
correspondance, so that it may be useful to look for their Bethe 
equations, their Bethe eigenvalues and vectors. To encompass future 
possible developements, we treat both the 
deformed and undeformed cases. We focus on distinguished gradation, 
as dealt in section \ref{sec:dist-gr}. The transfer matrix 
eigenvalues are then given by (\ref{eq:Lambda-dist}), where the weights 
$\Lambda_{j}(u)$ depends on representations at each site.
If we focus on fundamental representations on each site, 
with inhomogeneity parameters 
$a_{l}$, $l=1,\ldots,L$, they take the form
\begin{eqnarray}
\Lambda_{1}(u) &=& \begin{cases}\displaystyle
\prod_{l=1}^L (u-a_{l}-\hbar) \\[1.7ex]
\displaystyle
\prod_{l=1}^L (\frac{q\,u}{a_{l}}-\frac{a_{l}}{q\,u}) 
 \end{cases} \qquad
\Lambda_{j}(u) = \begin{cases}\displaystyle
\prod_{l=1}^L (u-a_{l}) \\[1.7ex]
\displaystyle
\prod_{l=1}^L (\frac{u}{a_{l}}-\frac{a_{l}}{u}) 
\end{cases}
\qquad j=2,\ldots,\fm+\fn-1\qquad
\end{eqnarray}
where the first line corresponds to $\cY(\fm|\fn)$ and the second one to 
$\cU_{q}(\fm|\fn)$.

\subsection{$\cA_{2|1}$ spin chains}
In addition to the $\cA_{2}$ Bethe vectors (\ref{eq:Bethegl2}), one 
can consider 
the vectors (\ref{eq:Bethegl21}) that simplifies as (up to a normalisation 
coefficient):
\begin{eqnarray}
\Phi^{3}_{M+1}(u^{(1)}_1,\dots,u^{(1)}_M,u^{(2)}_1) &=&
t_{12}(u^{(1)}_1) \cdots t_{12}(u^{(1)}_M)\,t_{23}(u_1^{(2)}) 
\,\Omega\label{eq:Bethegl21bis}
 \\
&-&\sum_{i=1}^M
\frac{\fc_{23}(u_1^{(2)},u^{(1)}_i)}{\fb(u^{(2)}_1,u_i^{(1)})}
\,\left(\prod_{k=i+1}^{M}\ff(u^{(1)}_i,u_1^{(2)})\right)
\nonu
&&\quad\times
t_{12}(u^{(1)}_1) \cdots t_{12}(u^{(1)}_{i-1})\,
t_{13}(u_i^{(1)})\,t_{12}(u^{(1)}_{i+1}) \cdots t_{12}(u^{(1)}_M)\,
t_{22}(u_1^{(2)})\,\Omega 
\nonumber
\end{eqnarray}
where the function $\ff(u,v)$ is given in (\ref{eq:deff}) and the 
functions $\fb(u,v)$ and $\fc_{jk}(u,v)$ are given in 
(\ref{eq:defac-Y})-(\ref{eq:defac-Uq}).
The form of the Bethe equations depend on the superalgebra one chooses:

\paragraph{$gl(2|1)$ spin chain}
\ben
\frac{\Lambda_{2} (u^{(1)}_j) }{\Lambda_1 (u^{(1)}_j) } &=&
-\,\prod_{i \neq j}^{M_1} 
\frac{u_i^{(1)}-u_j^{(1)}-\hbar}{u_i^{(1)}-u_j^{(1)}+\hbar}\
\prod_{i=1}^{M_{2}}   
\frac{u_i^{(2)}-u_j^{(1)}}{u_i^{(2)}-u_j^{(1)}-\hbar}
\qquad j=1,\ldots,M_{1}\\
\frac{\Lambda_{3} (u^{(2)}_j) }{\Lambda_2 (u^{(2)}_j) } &=&
-\,\prod_{i=1}^{M_{1}}    
\frac{u_j^{(2)}-u_i^{(1)}-\hbar}{u_j^{(2)}-u_i^{(1)}}
\qquad j=1,\ldots,M_{2}
\een

\paragraph{$\cU_{q}(gl(2|1))$ spin chain}
\begin{eqnarray*}
\frac{\Lambda_{2} (u^{(1)}_j) }{\Lambda_1 (u^{(1)}_j) } &=&
-\prod_{i \neq j}^{M_1} 
\frac{q\,(u_i^{(1)})^2-q^{-1}\,(u_j^{(1)})^2}
{q^{-1}\,(u_i^{(1)})^2-q\,(u_j^{(1)})^2}\ 
\prod_{i=1}^{M_{2}}   
\frac{(u_i^{(2)})^2-(u_j^{(1)})^2}
{q\,(u_i^{(2)})^2-q^{-1}\,(u_j^{(1)})^2}\ 
\qquad j=1,\ldots,M_{1}\,,\\
\frac{\Lambda_{3} (u^{(2)}_j) }{\Lambda_2 (u^{(2)}_j) } &=&
-\prod_{i=1}^{M_{1}}    
\frac{q\,(u_j^{(2)})^2-q^{-1}\,(u_i^{(1)})^2}
{(u_j^{(2)})^2-(u_i^{(1)})^2}
\qquad j=1,\ldots,M_{2}
\end{eqnarray*}

\subsection{$\cA_{2|2}$ spin chain}
In addition to the vectors (\ref{eq:Bethegl2}) and 
(\ref{eq:Bethegl21bis}), the vector (\ref{eq:Bethegl22}) rewrites
(up to normalisation):
\ben
\Phi^{4}_{M+2}(\{u\}) &=& 
 t_{12}(u_1^{(1)}) \dots t_{12}(u_M^{(1)})\,t_{23}(u_1^{(2)}) 
 \,t_{34}(u_1^{(3)})\, \Omega
 \label{eq:Bethegl22bis}\\
&+&
\frac{\fc_{34}(u_1^{(3)},u_1^{(2)})}{\fb(u_1^{(3)},u_1^{(2)})}
\ t_{12}(u_1^{(1)}) \dots t_{12}(u_M^{(1)})\,t_{24}(u_j^{(2)}) 
\,t_{33}(u_1^{(3)})\, \Omega
\nonumber \\
&-&\sum_{k=1}^{M_1} 
\frac{\fc_{23}(u_1^{(2)},u_i^{(1)})}{\fb(u_1^{(2)},u_i^{(1)})}
\left(\prod_{j=k+1}^{M_1}
\ff(u_j^{(1)},u_1^{(2)}) \right)
\nonumber \\
&&\qquad \times t_{12}(u_1^{(1)}) \dots  t_{12}(u_{k-1}^{(1)})
\,t_{13}(u_k^{(1)})\,t_{12}(u_{k+1}^{(1)})
\dots t_{12}(u_M^{(1)})\,
t_{22}(u_1^{(2)})\,t_{34}(u_1^{(3)})
\, \Omega
\nonumber \\
&-&\sum_{k=1}^{M_1}
\frac{\fc_{24}(u_1^{(2)},u_k^{(1)})}{\fb(u_1^{(2)},u_k^{(1)})}
\frac{\fc_{34}(u_1^{(3)},u_1^{(2)})}{\fb(u_1^{(3)},u_1^{(2)})}
\left(\prod_{j=k+1}^{M_1}
\ff(u_j^{(1)},u_1^{(2)})\right) 
\nonumber \\
&&\qquad\times   t_{12}(u_1^{(1)}) \dots t_{12}(u_{k-1}^{(1)})
\,t_{14}(u_k^{(1)})\,t_{12}(u_{k+1}^{(1)})
\dots t_{12}(u_M^{(1)})\,t_{22}(u_1^{(2)})\,
t_{33}(u_1^{(3)})\, \Omega
\nonumber
\een

\paragraph{$gl(2|2)$ spin chain}
\ben
\frac{\Lambda_{2} (u^{(1)}_j) }{\Lambda_1 (u^{(1)}_j) } &=&
-\,\prod_{i \neq j}^{M_1} 
\frac{u_i^{(1)}-u_j^{(1)}-\hbar}{u_i^{(1)}-u_j^{(1)}+\hbar}\
\prod_{i=1}^{M_{2}}   
\frac{u_i^{(2)}-u_j^{(1)}}{u_i^{(2)}-u_j^{(1)}-\hbar}
\qquad j=1,\ldots,M_{1}
\\
\frac{\Lambda_{3} (u^{(2)}_j) }{\Lambda_2 (u^{(2)}_j) } &=&
-\,\prod_{i=1}^{M_{1}}    
\frac{u_j^{(2)}-u_i^{(1)}-\hbar}{u_j^{(2)}-u_i^{(1)}}\ 
\prod_{i=1}^{M_{3}}   
\frac{u_i^{(3)}-u_j^{(2)}}{u_i^{(3)}-u_j^{(2)}+\hbar}
\qquad j=1,\ldots,M_{2}
\\
\frac{\Lambda_{4} (u^{(3)}_j) }{\Lambda_3 (u^{(3)}_j) } &=&
-\,\prod_{i=1}^{M_{2}}    
\frac{u_j^{(3)}-u_i^{(2)}-\hbar}{u_j^{(3)}-u_i^{(2)}}\ 
\prod_{i \neq j}^{M_{3}}   
\frac{u_i^{(3)}-u_j^{(3)}+\hbar}{u_i^{(3)}-u_j^{(3)}-\hbar}
\qquad j=1,\ldots,M_{3}
\een

\paragraph{$\cU_{q}(gl(2|2))$ spin chain}
\begin{eqnarray*}
\frac{\Lambda_{2} (u^{(1)}_j) }{\Lambda_1 (u^{(1)}_j) } &=&
-\prod_{i \neq j}^{M_1} 
\frac{q\,(u_i^{(1)})^2-q^{-1}\,(u_j^{(1)})^2}
{q^{-1}\,(u_i^{(1)})^2-q\,(u_j^{(1)})^2}\ 
\prod_{i=1}^{M_{2}}   
\frac{(u_i^{(2)})^2-(u_j^{(1)})^2}
{q\,(u_i^{(2)})^2-q^{-1}\,(u_j^{(1)})^2}\ 
\qquad j=1,\ldots,M_{1}\,,
\\
\frac{\Lambda_{3} (u^{(2)}_j) }{\Lambda_2 (u^{(2)}_j) } &=&
-\prod_{i=1}^{M_{1}}    
\frac{q\,(u_j^{(2)})^2-q^{-1}\,(u_i^{(1)})^2}
{(u_j^{(2)})^2-(u_i^{(1)})^2}\ 
\prod_{i=1}^{M_{3}}   
\frac{(u_i^{(3)})^2-(u_j^{(2)})^2}
{q^{-1}\,(u_i^{(3)})^2-q\,(u_j^{(2)})^2}
\qquad j=1,\ldots,M_{2}
\\
\frac{\Lambda_{4} (u^{(3)}_j) }{\Lambda_3 (u^{(3)}_j) } &=&
-\prod_{i=1}^{M_{2}}    
\frac{q\,(u_j^{(3)})^2-q^{-1}\,(u_i^{(2)})^2}
{(u_j^{(3)})^2-(u_i^{(2)})^2}\ 
\prod_{i \neq j}^{M_{3}}   
\frac{q\,(u_i^{(3)})^2-q^{-1}\,(u_j^{(3)})^2}
{q^{-1}\,(u_i^{(3)})^2-q\,(u_j^{(3)})^2}
\qquad j=1,\ldots,M_{3}
\end{eqnarray*}

\subsection{$\cA_{4|4}$ spin chain}
The form of new Bethe vectors is 
becoming very complicated and we refrain from giving an example. 
However, since $\cA_{4|4}$ is the first superalgebra in the 
`super-Yang-Mills series' 
$\cA_{2}\equiv\cA_{2|0}$, $\cA_{2|1}$, $\cA_{2|2}$, $\cA_{4|4}$, that leads to 
generic Bethe ansatz equations, we write them:

\paragraph{$gl(4|4)$ spin chain}
\begin{eqnarray*}
\frac{\Lambda_{2} (u^{(1)}_j) }{\Lambda_1 (u^{(1)}_j) } &=&
-\,\prod_{i \neq j}^{M_1} 
\frac{u_i^{(1)}-u_j^{(1)}-\hbar}{u_i^{(1)}-u_j^{(1)}+\hbar}\
\prod_{i=1}^{M_{2}}   
\frac{u_i^{(2)}-u_j^{(1)}}{u_i^{(2)}-u_j^{(1)}-\hbar}
\qquad j=1,\ldots,M_{1}
\\
\frac{\Lambda_{k+1} (u^{(k)}_j) }{\Lambda_k (u^{(k)}_j) } &=&
-\,\prod_{i=1}^{M_{k-1}}    
\frac{u_j^{(k)}-u_i^{(k-1)}-\hbar}{u_j^{(k)}-u_i^{(k-1)}}\ 
\prod_{i \neq j}^{M_k} 
\frac{u_i^{(k)}-u_j^{(k)}-\hbar}{u_i^{(k)}-u_j^{(k)}+\hbar}\
\prod_{i=1}^{M_{k+1}}   
\frac{u_i^{(k+1)}-u_j^{(k)}}{u_i^{(k+1)}-u_j^{(k)}-\hbar}
\nonu
j&=&1,\ldots,M_{k}\,,\quad k=2,\,3
\end{eqnarray*}
\begin{eqnarray*}
\frac{\Lambda_{5} (u^{(4)}_j) }{\Lambda_4 (u^{(4)}_j) } &=&
-\,\prod_{i=1}^{M_{3}}    
\frac{u_j^{(4)}-u_i^{(3)}-\hbar}{u_j^{(4)}-u_i^{(3)}}\ 
\prod_{i=1}^{M_{5}}   
\frac{u_i^{(5)}-u_j^{(4)}}{u_i^{(5)}-u_j^{(4)}+\hbar}
\qquad j=1,\ldots,M_{4}
\end{eqnarray*}
\begin{eqnarray*}
\frac{\Lambda_{k+1} (u^{(k)}_j) }{\Lambda_k (u^{(k)}_j) } &=&
-\,\prod_{i=1}^{M_{k-1}}    
\frac{u_j^{(k)}-u_i^{(k-1)}+\hbar}{u_j^{(k)}-u_i^{(k-1)}}\ 
\prod_{i \neq j}^{M_k} 
\frac{u_i^{(k)}-u_j^{(k)}+\hbar}{u_i^{(k)}-u_j^{(k)}-\hbar}\
\prod_{i=1}^{M_{k+1}}   
\frac{u_i^{(k+1)}-u_j^{(k)}}{u_i^{(k+1)}-u_j^{(k)}+\hbar}
\nonu
j&=&1,\ldots,M_{k}\,,\quad k=5,6
\\[1.4ex]
\frac{\Lambda_{8} (u^{(7)}_j) }{\Lambda_k (u^{(7)}_j) } &=&
-\,\prod_{i=1}^{M_{6}}    
\frac{u_j^{(7)}-u_i^{(6)}+\hbar}{u_j^{(7)}-u_i^{(6)}}\ 
\prod_{i \neq j}^{M_7} 
\frac{u_i^{(7)}-u_j^{(7)}+\hbar}{u_i^{(7)}-u_j^{(7)}-\hbar}
\qquad j=1,\ldots,M_{7}\,,
\end{eqnarray*}

\paragraph{$\cU_{q}(gl(4|4))$ spin chain}
\begin{eqnarray*}
\frac{\Lambda_{2} (u^{(1)}_j) }{\Lambda_1 (u^{(1)}_j) } &=&
-\prod_{i \neq j}^{M_1} 
\frac{q\,(u_i^{(1)})^2-q^{-1}\,(u_j^{(1)})^2}
{q^{-1}\,(u_i^{(1)})^2-q\,(u_j^{(1)})^2}\ 
\prod_{i=1}^{M_{2}}   
\frac{(u_i^{(2)})^2-(u_j^{(1)})^2}
{q\,(u_i^{(2)})^2-q^{-1}\,(u_j^{(1)})^2}\ 
\qquad j=1,\ldots,M_{1}\,,
\\
\frac{\Lambda_{k+1} (u^{(k)}_j) }{\Lambda_k (u^{(k)}_j) } &=&
-\prod_{i=1}^{M_{k-1}}    
\frac{q\,(u_j^{(k)})^2-q^{-1}\,(u_i^{(k-1)})^2}
{(u_j^{(k+1)})^2-(u_i^{(k)})^2}\ 
\prod_{i \neq j}^{M_k} 
\frac{q\,(u_i^{(k)})^2-q^{-1}\,(u_j^{(k)})^2}
{q^{-1}\,(u_i^{(k)})^2-q\,(u_j^{(k)})^2}\\
&&\times \prod_{i=1}^{M_{k+1}}   
\frac{(u_i^{(k+1)})^2-(u_j^{(k)})^2}
{q\,(u_i^{(k+1)})^2-q^{-1}\,(u_j^{(k)})^2}
\qquad j= 1,\ldots,M_{k}\,,\qquad k=2,\,3
\end{eqnarray*}
\begin{eqnarray*}
\frac{\Lambda_{5} (u^{(4)}_j) }{\Lambda_4 (u^{(4)}_j) } &=&
-\prod_{i=1}^{M_{3}}    
\frac{q\,(u_j^{(4)})^2-q^{-1}\,(u_i^{(3)})^2}
{(u_j^{(4)})^2-(u_i^{(3)})^2}\ 
\prod_{i=1}^{M_{5}}   
\frac{(u_i^{(5)})^2-(u_j^{(4)})^2}
{q^{-1}\,(u_i^{(5)})^2-q\,(u_j^{(4)})^2}
\qquad j=1,\ldots,M_{4}
\end{eqnarray*}
\begin{eqnarray*}
\frac{\Lambda_{k+1} (u^{(k)}_j) }{\Lambda_k (u^{(k)}_j) } &=&
-\prod_{i=1}^{M_{k-1}}    
\frac{q^{-1}\,(u_j^{(k)})^2-q\,(u_i^{(k-1)})^2}
{(u_j^{(k)})^2-(u_i^{(k-1)})^2}\ 
\prod_{i \neq j}^{M_k} 
\frac{q^{-1}\,(u_i^{(k)})^2-q\,(u_j^{(k)})^2}
{q\,(u_i^{(k)})^2-q^{-1}\,(u_j^{(k)})^2}\\
&&\times \prod_{i=1}^{M_{k+1}}   
\frac{(u_i^{(k+1)})^2-(u_j^{(k)})^2}
{q^{-1}\,(u_i^{(k+1)})^2-q\,(u_j^{(k)})^2}
\qquad j= 1,\ldots,M_{k}\,,\quad k=5,6
\\[1.4ex]
\frac{\Lambda_{8} (u^{(7)}_j) }{\Lambda_7 (u^{(7)}_j) } &=&
-\prod_{i=1}^{M_{6}}    
\frac{q^{-1}\,(u_j^{(7)})^2-q\,(u_i^{(6)})^2}
{(u_j^{(7)})^2-(u_i^{(6)})^2}\ 
\prod_{i \neq j}^{M_7} 
\frac{q^{-1}\,(u_i^{(7)})^2-q\,(u_j^{(6)})^2}
{q\,(u_i^{(7)})^2-q^{-1}\,(u_j^{(6)})^2}\ 
\qquad j=1,\ldots,M_{7}\,,
\end{eqnarray*}

\subsection*{Acknowledgments}
Sections \ref{sect:indec} and \ref{sect:ex-super} follow from useful  
 comments of the referee: we thank him for his remarks. \\
We also thank Paul Sorba for quoting references \cite{vlad} 
and \cite{marcu}.


\appendix

\section{Finite dimensional algebras}
\subsection{$gl(\fn)$ and $gl(\fm|\fn)$ algebras\label{app.A}}

The Lie algebra $gl(\fn)$ is  a vector space over $\CC$ spanned by the 
generators $\{\cE_{ij} | i,j=1,2,...,\fn 
\}$. 
The bilinear commutator associated to $gl(\fn)$ is defined by: 
\begin{equation}
[ \cE_{ij},\cE_{kl}]
=\delta_{kj}\,\cE_{il}-\delta_{il}\,\cE_{kj}\,.
\end{equation}

The Lie superalgebra $gl(\fm|\fn)$  is a  $\ZZ_2$-graded vector space 
over $\CC$ spanned by the generators $\{\cE_{ij}| i,j=1,2,...,\fm+\fn\}$. 
The bilinear graded commutator 
associated to $gl(\fm|\fn)$ is defined by: 
\begin{equation}
{[}\cE_{ij},\,\cE_{kl}\}
=\delta_{kj}\,\cE_{il}-(-1)^{([i]+[j])([k]+[l])}
\delta_{il}\,\cE_{kj}\,.
\end{equation}
It is graded anti-symmetric:
\begin{equation}
{[}\cE_{ij},\,\cE_{kl}\}=-(-1)^{([i]+[j])([k]+[l])}\,{[}\cE_{kl},\,\cE_{ij}\}
\end{equation}

The highest weight representations of these Lie (super)algebras are 
characterized by the highest weight 
$\lambda = (\lambda_1, \dots, \lambda_{\fm+\fn})$ (with $\fm=0$ for the 
non-graded case). Finite dimensional irreducible representations are 
obtained when the following relations are obeyed:
\begin{eqnarray}
&& \lambda_{i}-\lambda_{i+1}\in\ZZ_{+}\,,\ i=1,2,\ldots,\fn-1 
\mb{for} gl(\fn) \label{eq:hw-gl}\\
&& 
\begin{cases}
\lambda_{i}-\lambda_{i+1}\in\ZZ_{+}\,,\ i=1,2,\ldots,\fm-1\\
\lambda_{i}-\lambda_{i+1}\in\ZZ_{+}\,,\ i=\fm+1,\ldots,\fn+\fm-1
\end{cases} \mb{for} gl(\fm|\fn)
\label{eq:hw-glmn}
\end{eqnarray}
 
\subsection{$\cU_q(gl(\fn))$ and $\cU_q(gl(\fm|\fn))$ algebra\label{app.B}}
We suppose that $q$ is not a root of unity.

$\cU_q(gl(\fn))$ is an associative algebra over $\CC$ generated by 
$q^{\pm H_j}$, $e_i$ and $f_i$ ($1\leq j\leq\fn$,
$1\leq i\leq\fn-1$) with the relations:
\ben
&&q^{H_i}\,q^{-H_i}=q^{-H_i}\,q^{H_i}=1\\ 
&&q^{H_i}\,e_j\,q^{-H_i}=q^{\delta_{ij}-\delta_{i j+1}}\,e_j\\ 
&&q^{H_i}\,f_j\,q^{-H_i}=q^{-\delta_{ij}+\delta_{i j+1}}\,f_j\\ 
&&e_i\,f_j-f_j\,e_i=\delta_{ij} \frac{ q^{H_i-H_{i+1}}-q^{-H_i+H_{i+1}} 
}{q-q^{-1}}\\ 
&&e_i\,e_j=e_j\,e_i \mb{,} f_i\,f_j=f_j\,f_i  \mb{for}|i-j|\geq 2\\ 
&&e_i^2\,e_{i\pm1}-(q-q^{-1})\,e_i\,e_{i\pm1}\,e_i+e_{i\pm1}\,e_i^2=0
\mb{for}1\leq i\mb{and}i\pm1\leq \fn\\ 
&&f_i^2\,f_{i\pm1}-(q-q^{-1})\,f_i\,f_{i\pm1}\,f_i+f_{i\pm1}\,f_i^2=0
 \mb{for}1\leq i\mb{and} i\pm1\leq\fn
\qquad\qquad
\een
The highest weight finite-dimensional irreducible
representations of $\cU_{q}(gl(\fn))$, are characterized by a $gl(\fn)$ highest
weight $\lambda = (\lambda_1,\dots,\lambda_\fn)$ as given in 
(\ref{eq:hw-gl}) and a set of parameters $\eta_{j}=\pm1,\pm i$ 
$(j=1,\ldots,\fn)$, see \cite{Rosso} for more details. 
Explicitly, 
the weights of the $\cU_{q}(\fn)$ algebra read:
\begin{equation}
(\eta_{1}\,q^{\lambda_{1}},\eta_{2}\,q^{\lambda_{2}},
\ldots,\eta_{\fn}\,q^{\lambda_{\fn}}) \mb{with} 
\lambda_{j}-\lambda_{j+1}\in\ZZ_{+}
\mb{and} \eta_{j}=\pm1,\pm i
\end{equation}

$\cU_q(gl(\fm|\fn))$ is an associative algebra over $\CC$ generated by 
$q^{\pm H_j}$, $e_i$ and $f_i$ ($1\leq j\leq\fm+\fn$,
$1\leq i\leq \fm+\fn-1$) with the defining relations:
\ben
&&q^{H_i}\,q^{-H_i}=q^{-H_i}\,q^{H_i}=1\\ 
&&q^{H_i}\,e_j\,q^{-H_i}=q^{(-1)^{[i]}(\delta_{ij}-\delta_{i j+1})}\,e_j\\ 
&&q^{H_i}\,f_j\,q^{-H_i}=q^{-(-1)^{[i]}(\delta_{ij}-\delta_{i j+1})}
\,f_j
\een
\ben
&&e_i\,f_j-(-1)^{([i]+[i+1])([j]+[j+1])}\,f_j\,e_i=\delta_{ij} \frac{ 
q^{H_i-H_{i+1}}-q^{-H_i+H_{i+1}} }{q^{(-1)^{[i]}}-q^{-(-1)^{[i]}}}\\ 
&&e_i\,e_j=(-1)^{([i]+[i+1])([j]+[j+1]}\,e_j\,e_i \mb{,} 
f_i\,f_j=(-1)^{([i]+[i+1])([j]+[j+1]}\,f_j\,f_i  
\\ 
&&e_i^2\,e_{i\pm1}-(q-q^{-1})\,e_i\,e_{i\pm1}\,e_i+e_{i\pm1}\,e_i^2=0 
\mb{with} i \neq \fm\\ 
&&f_i^2\,f_{i\pm1}-(q-q^{-1})\,f_i\,f_{i\pm1}\,f_i+f_{i\pm1}\,f_i^2=0 
\mb{with} i \neq \fm
\een

The following identification gives the isomorphism between the RTT 
presentation and 
the Serre-Chevalley one \cite{FRT}.
\ben
l^{+}_{ii}&=&(-1)^{[i]}\,q^{H_i}\mb{;}
l^+_{i,i+1}=(-1)^{[i+1]}\,(q-q^{-1})\,q^{H_i}\,f_i\mb{;}
l^+_{i,i-1}=0\\ 
l^{-}_{ii}&=&(-1)^{[i]}\,q^{- H_i}\mb{;}
l^-_{i,i-1}=(-1)^{[i-1]}\,(q-q^{-1})\,e_i\,q^{-H_i}\mb{;}
l^-_{i,i+1}=0
\een

Highest weight finite-dimensional irreducible
representations of $\cU_{q}(gl(\fm|\fn))$ have been studied in 
\cite{Z2}. They are characterized by a $gl(\fm|\fn)$ highest
weight $\lambda = (\lambda_1,\dots,\lambda_{\fm+\fn})$ as given in 
(\ref{eq:hw-glmn}) and a set of parameters $\eta_{a}$: 
\begin{equation}
((-1)^{[1]}\eta_{1}\,q^{\lambda_{1}},
\ldots,(-1)^{[\fm+\fn]}\eta_{\fm+\fn}\,q^{\lambda_{\fm+\fn}}) \mb{with} 
\begin{cases}
\lambda_{j}-\lambda_{j+1}\in\ZZ_{+}\,,\ j\neq\fm\\
\eta_{j}=\pm1\,,\ j=1,\ldots,\fn+\fm
\end{cases}
\end{equation}


\end{document}